\documentclass{LMCS}

\def\doi{8 (1:32) 2012}                                                  
\lmcsheading%
{\doi}                                                                      
{1--32}                                                                    
{}                                                                         
{}                                                                        
{
Mar.~\phantom02, 2010}                                                    
{Mar.~29, 2012}                                                            
{}                                                                              

\usepackage{amsmath, amssymb}
\usepackage[british]{babel}
\usepackage{tikz}
\usetikzlibrary{snakes}
\usepackage{enumerate,hyperref}

\newcommand{\ncm}[1]{\expandafter\newcommand\csname#1\endcsname}
\ncm{calletter}[1]{\ncm{x#1}{\mathcal{#1}}}
\calletter{C}
\calletter{I}
\ncm{frakletter}[1]{\ncm{f#1}{\mathfrak{#1}}}
\frakletter A
\ncm{makemathname}[1]{\mathrm{#1}}
\ncm{mathname}[2]{\ncm{#1}{\makemathname{#2}}}
\ncm{mathownname}[1]{\mathname{#1}{#1}}
\ncm{vfi}{\lessgtr_{\phi}}

\newtheorem{fct}[thm]{Fact}
\newtheorem{clam}{Claim}
\ncm{OO}[1]{O(#1)}
\mathname{dd}{ead}
\mathownname{var}
\mathownname{free}
\mathownname{tw}
\mathownname{stw}
\mathownname{fotw}
\mathownname{w}
\mathownname{ew}
\mathownname{ar}
\mathownname{cw}
\mathname{moncw}{mon\text-cw}
\mathownname{poly}
\mathname{Int}{int}
\mathownname{ext}
\newcommand{\bigmid}{\;\big|\;}

\ncm{cqed}{\hspace*{\fill}${}_\blacksquare$}

\ncm{ml}{\mathcal L}
\ncm{foks}[1]{\ml^{#1}}
\ncm{fokm}[1]{\tilde{\ml}^{#1}}
\ncm{fok}{\foks k}
\newcommand{\gfi}{G_{\varphi}}

\newcommand{\nrsd}{\textsc{nrsd}}
\renewcommand{\phi}{\varphi}
\newcommand{\ad}{\dd_{\varphi}}

\newcommand{\fpt}{\textsc{fpt}}
\newcommand{\mc}{\textsc{MC}}
\newcommand{\cq}{\textsc{CQ}}
\newcommand{\eval}{\textsc{Eval}}
\newcommand{\mM}{\mathcal A}
\newcommand{\rahmen}[1]{\begin{center}\fbox{\parbox{7.5cm}{#1}}\end{center}}
\newcommand{\linie}{\\[-2mm]\hrule\mbox{}\\}

\newcommand{\refl}{Reflexivity}
\newcommand{\trans}{Transitivity}
\newcommand{\alt}{Alternation}
\newcommand{\alte}{\makemathname{ad}}
\ncm{altfi}{\alte_{\phi}}

\begin{document}\setlength{\labelwidth}{5cm}

\title{Tree-width for first order formulae}

\author[I.~Adler]{Isolde Adler\rsuper a}
\address{{\lsuper a}Institut f\"ur Informatik, Goethe-Universit\"at Frankfurt am Main}
\email{iadler@informatik.uni-frankfurt.de}

\author[M.~Weyer]{Mark Weyer\rsuper b}
\address{{\lsuper b}not affiliated}
\email{mark@weyer-zuhause.de}


\begin{abstract} 
We introduce tree-width for first order formulae $\varphi$, $\fotw(\varphi)$.
We show that computing $\fotw$ 
is fixed-parameter tractable with parameter $\fotw$.
Moreover, we show that on classes of
formulae of bounded $\fotw$, model checking is fixed parameter tractable, with
parameter the length of the formula. This is done by translating a formula
$\varphi$ with $\fotw(\varphi)<k$ into a formula of the $k$-variable
fragment $\fok$ of first order logic. For fixed $k$, 
the question whether a given first order formula is equivalent 
to an $\fok$ formula is undecidable. In contrast, 
the classes of first order formulae with bounded $\fotw$ are
fragments of first order logic for which the equivalence is decidable.

Our notion of tree-width generalises tree-width of conjunctive queries to
arbitrary formulae of first order logic by taking into account the quantifier
interaction in a formula. Moreover, it is more powerful than the notion of
elimination-width 
of quantified constraint formulae, defined by Chen and Dalmau 
(CSL 2005):
for quantified constraint formulae, both bounded elimination-width
and bounded $\fotw$ allow for model checking in polynomial time.
We prove that $\fotw$ of a quantified constraint formula $\phi$ is
bounded by the elimination-width of $\phi$, and
we exhibit a class of quantified constraint formulae
with bounded $\fotw$, that has unbounded elimination-width.
A similar comparison holds for strict tree-width of non-recursive stratified datalog 
as defined by Flum, Frick, and Grohe (JACM 49, 2002). 

Finally, we show that 
$\fotw$ has a characterization in terms of 
a cops and robbers game without monotonicity cost.
\end{abstract}

\maketitle

\section{Introduction}

Model checking is an important problem in complexity theory. It asks
for a given formula $\phi$ of some class $\mathcal C$ of formulae 
and a structure $\mathcal A$,
whether $\mathcal A$ satisfies $\phi$.

\smallskip\noindent\rahmen
{
	$\mc(\mathcal C)$ 
	\linie
	\textbf{Input:} A structure $\mM$ and a formula $\varphi\in\mathcal C$.\\
	\textbf{Question:} $\mM\models\phi$?
}

Let $\ml$ denote first order logic.
It is well-known, that $\mc(\ml)$ is \textsc{pspace}-complete.
Motivated by this, much research has been done on finding 
fragments of $\ml$  having a tractable model checking problem.
For instance, for fixed $k$, the problem $\mc(\fok)$ can be solved 
in polynomial time, 
where $\fok$ denotes the fragment of first order formulae with at most
$k$ variables (see e.g.~\cite{FG06}).

The class of \emph{conjunctive queries}, $\cq$, is an important fragment of first order logic.
Many queries that occur in practice are conjunctive queries, and
model checking of conjunctive queries on relational databases 
(i.e.~relational structures) is an important 
and well-studied problem in database theory \cite{yan81,cheraj00,fre90,dalkolvar02,gotleosca02,gromar06}.
It is equivalent to conjunctive query containment, to
the constraint satisfaction problem studied in artificial intelligence
and to the homomorphism problem for structures \cite{chamer77,fedvar98}.
A \emph{conjunctive query} is a first order formula starting with a quantifier prefix using
only existential quantifiers, followed by a conjunction of relational atoms.
While $\mc(\cq)$ is NP-hard in general, 
several researchers proved independently
that conjunctive queries of bounded \emph{tree-width} can be 
evaluated in polynomial time \cite{cheraj00,fre90}. 
One way to prove this is the following.
Suppose $\phi$ is a conjunctive query having tree-width $k$. 
Then we can compute a tree decomposition of width $k$ in linear time
using Bodlaender's algorithm \cite{Bodlaender96}. {}From the decomposition
we can actually read off the syntax of an equivalent formula 
$\phi'\in\foks{k+1}$.
Finally, we use the fact that $\mc(\foks{k+1})$ 
is solvable in polynomial time.
Essentially, bounded tree-width is even necessary for polynomial time
solvability of $\mc(\cq)$~\cite{GroheSS01,Grohe07}. 

In this paper, we introduce a notion of \emph{tree-width for first order formulae}
$\phi$, $\fotw(\phi)$. Our notion generalises the notion of tree-width of conjunctive queries, and
we show that the class $\mathcal C_k$ of all first order formulae $\phi$ with $\fotw(\phi)\leq k$
satisfies the following properties.
\begin{enumerate}[(1)]
\item $\mathcal C_k$ has a polynomial time membership test 
		(Corollary~\ref{cor:computing-decompositions}).
\item $\mathcal C_k$ has the same expressive power as $\foks{k+1}$, 
		the fragment of
		first order formulae with at most $k+1$ 
		variables (Theorem \ref{theo:fotw-fok}). 
	\item There is an algorithm that computes for given $\phi\in\mathcal C_k$
		an equivalent formula $\phi'\equiv \phi$ with $\phi'\in\foks{k+1}$ 
		(Theorem \ref{theo:fotw-fok}).
	\item $\mc(\mathcal C_k)$ is fixed parameter tractable 
		with parameter the length of $\phi$, i.e.\ for input $\phi\in\mathcal C_k$
		and $\mathcal A$, the running time is $p(\|\mathcal A\|)f(\left|\phi\right|)$
		for a polynomial $p$ and a computable function $f$ (Corollary \ref{cor:fo-eval}).
\end{enumerate}
Obviously, properties 1 and 3 imply property 4.
While $\mc(\fok)$ is solvable in polynomial time, 
we do not obtain a polynomial algorithm for $\mc(\mathcal C_k)$. 
Nevertheless, in typical applications one can expect the length
of the formula to be small compared to the size of the structure (database).
For a fixed formula the running time is polynomial, and moreover, the problem
is \emph{fixed-parameter tractable} (in \fpt), 
meaning that changing $\phi$ 
does not alter the exponent of the polynomial (see \cite{DF99,FG06}).

Note that for fixed $k>0$ it is undecidable, whether a first order formula
$\phi$ is equivalent to an $\fok$ formula.
Hence it is not surprising that our notion of $k$-bounded first order tree-width does not capture 
semantic equivalence to $\fok$ (we will give more details in Section~\ref{section:mc}). 

\emph{Quantified constraint formulae} generalise conjunctive queries by allowing
arbitrary quantifiers in the quantifier prefix.  In \cite{chedal05}, Chen
and Dalmau introduce \emph{elimination orderings} for quantified
constraint formulae. 
These elimination orderings must respect the quantifier prefix.
In this way, Chen and Dalmau obtain a notion of \emph{elimination-width}\footnote{Actually, 
the notion is called \emph{tree-width} for quantified constraint formulae in \cite{chedal05}, 
But since the notion is defined via elimination orderings, we prefer the term
\emph{elimination-width}.},
which allows for model checking of quantified constraint formulae of bounded 
elimination-width in polynomial time, using a consistency algorithm. Hereby, they
answer a question posed in \cite{GottlobGS05} positively, whether bounded tree-width methods
work for formulae more general than conjunctive queries. 
Introducing a notion of tree-width 
for arbitrary first order formulae, we even go further.  
We show that for quantified constraint formulae $\phi$, 
elimination-width of $\phi$ is at least as large as $\fotw(\phi)$,
and we exhibit a class of quantified constraint formulae
with bounded first order tree-width and unbounded elimination-width.
We show that quantified constraint formulae of bounded $\fotw$ allow for 
model checking in polynomial time. 
Hence $\fotw$ is more powerful than elimination-width. 

In \cite{flufrigro01}, Flum, Frick and Grohe introduce 
\emph{strict tree-width}\footnote{In \cite{flufrigro01}, 
the authors also introduce a notion of \emph{tree-width for first order formulae}. But their notion disregards the quantifier 
interaction, and they only use it
for conjunctive queries with negation.}
for \emph{non-recursive stratified datalog ($\nrsd$) programs}. They show that
model checking for $\nrsd$ programs of bounded strict tree-width can be done
in polynomial time. Since $\nrsd$ programs have a canonical translation
into first order formulae, 
our notion of tree-width can be transfered 
from first order formulae to $\nrsd$ programs. 
We show that if an $\nrsd$ program $\Pi$ 
has strict tree-width at most $k$, then the formula $\phi_{\Pi}$ obtained from
$\Pi$ has elimination-width at most $k$ and hence it 
satisfies $\fotw(\phi_{\Pi})\leq k$. 
Again there are classes of $\nrsd$ programs with unbounded strict tree-width, 
whose corresponding first order formulae have bounded first order 
tree-width.  Hence our notion of
first order tree-width yields larger subclasses of $\ml$, that still allow
for tractable model checking.

Actually, we introduce first order tree-width 
as a special case of a more abstract notion 
which we term stratified tree-width. 
We expect that stratified tree-width 
will find further, quite different, applications. 

The rest of this paper is organised as follows. 
Section~\ref{sec:prems} fixes some terminology. 
Section~\ref{section:fotw} introduces the notion 
of stratified tree-width, the special case of first order tree-width, 
and the notion of \emph{xenerp normal form} of a formula $\phi$ -- 
a kind of opposite of prenex normal form. 
We show that $\fotw$ is invariant under transformation into xenerp normal form.
In Section~\ref{section:comparing} we relate $\fotw$ to the
natural notion of tree-width stratified by 
the alternation depth of a formula.
In Section~\ref{sec:computing} 
we show how to compute stratified tree decompositions 
and, in particular, how to compute first order tree-width. 
In Section~\ref{section:mc} we prove 
that bounded first order tree-width 
is expressively equivalent to bounded variable fragments of first order logic
and that model checking for formulae of bounded first order tree-width 
is fixed-parameter tractable. 
In Section~\ref{section:related-notions} 
we relate our notion to existing notions 
and give a game characterisation of stratified tree-width. 
We conclude with some open problems in Section~\ref{section:conclusion}. 

We wish to thank the anonymous referees for many useful suggestions.

\section{Well-known definitions}\label{sec:prems}

A \emph{vocabulary} $\sigma=\{R_1,\ldots, R_n, c_1,\ldots,c_m\}$
is a finite set of \emph{relation symbols $R_i$}, $1\le i\le n$, and
\emph{constant symbols} $c_j$, $1\le j\le m$. 
Every $R_i$ has an associated \emph{arity}, an integer $\ar(R_i)> 0$.
A \emph{$\sigma$-structure} is a tuple 
$\mM=(A,R^{\mM}_1,\ldots, R^{\mM}_n,c^{\mM}_1,\ldots, c^{\mM}_m)$ where $A$ is
a finite set, the \emph{universe} of $\mM$, 
$R^{\mM}_i\subseteq A^{\ar(R_i)}$ for $1\leq i\leq n$, and
$c^{\mM}_j\in A$ for $1\leq j\leq m$.

Given a $\sigma$-structure $\mM$ we distinguish between the cardinality
$\left|A\right|$ of the universe $A$ of $\mM$ 
and the \emph{size} $\|\mM\|$ of $\mM$, given by 
$
	\|\mM\|=\left|\sigma\right|+\left|A\right|+
	\sum_{i=1}^n\left|R_i^{\mM}\right|\cdot\ar(R_i).
$ 

We use $\ml$ to denote relational first order logic with constants, and for simplicity,
we refer to $\ml$ as \emph{first order logic}.
We assume that the reader is familiar with the basic notions of first order logic
(see for instance \cite{ebbflu90}).
For a formula $\phi$ we let $\free(\phi)$ denote the set of free variables of $\phi$.
A formula $\phi$ is a \emph{sentence}, if $\free(\phi)=\emptyset$.
We sometimes write $\phi(x_1,\ldots,x_n)$ to indicate that 
$\free(\phi)\subseteq\{x_1,\ldots,x_n\}$. 

For a structure $\mathcal A$, a formula $\phi(x_1,\ldots,x_n)$,
and elements $a_1,\ldots a_n\in A$ we write
$\mathcal A\models \phi(a_1,\ldots a_n)$ to denote that $\mathcal A$ satisfies
$\phi$ if the variables $x_1,\ldots,x_n$ are interpreted by $a_1,\ldots, a_n$,
respectively.
We let
\[
	\phi(\mathcal A):=\{(a_1,\ldots, a_n)\mid \mathcal A\models \phi(a_1,\ldots, a_n)\}.
\]
For sentences we have $\phi(\mathcal A)=\textsc{true}$, if $\mathcal A$ satisfies 
$\phi$, and \textsc{false} otherwise. If the vocabularies of $\phi$ and
$\mathcal A$ are different, we let $\phi(\mathcal A)=\emptyset$.

The \emph{Query Evaluation Problem} for a class $\mathcal C$ of formulae is the following
problem:

\smallskip\noindent\rahmen
{
	$\eval(\mathcal C)$ 
	\linie
	\textbf{Input:} A structure $\mM$ and a formula $\varphi\in \mathcal C$.\\
	\textbf{Problem:} Compute $\phi(\mM)$.
}
Note that if $\phi$ is a sentence, then $\eval(\mathcal C)$ and $\mc(\mathcal C)$
coincide.
We say that a formula $\varphi\in\ml$ is \emph{straight}, 
if no variable in $\varphi$ is quantified over twice, 
if no free variable is also a quantified variable, and if
each quantified variable actually occurs in some atom. 
All formulae are straight, unless stated otherwise. 
Moreover, we assume that all formulae are in negation normal form, i.e.\ the
negation symbols only appear in front of atoms. 

We denote a graph $G$ as a pair $G=(V(G),E(G))$, where the set $V(G)$ of \emph{vertices}
is finite, and every \emph{edge} $e\in E(G)$ is a two-element subset of $V(G)$.
A \emph{tree decomposition} of a graph $G=(V,E)$ is a pair $(T,B)$, consisting
of a rooted tree $T$ and a family $B=(B_t)_{t \in T}$ of subsets of $V$, the
\emph{pieces} of $T$, satisfying:
	\begin{description}

	\item[(TD1)]
	For each $v \in V$ there exists $t \in T$, such that $v \in B_t$.
	We say the node $t$ \emph{covers}~$v$.

	\item[(TD2)]
	For each edge $e \in E$ there exists $t \in T$, such that $e \subseteq B_t$.
	We say the node $t$ \emph{covers} $e$.

	\item[(TD3)] For each $v \in V$ the set
	$\{t \in T \mid v \in B_t \}$ is connected in $T$.
	\end{description}

\noindent 
The \emph{width} of $(T,B)$ is defined as $\w(T,B):=
\max\big\{\left|B_t\right|\bigmid t\in T\big\}-1$. 

The \emph{tree-width of $G$} is defined as
\[
	\tw(G):= 
	\min\big\{\w(T,B)\bigmid (T,B) \textsl{ is a tree decomposition of }G\big\}.
\]

\begin{fct}\label{fact:number-of-edges}
	Every graph $G$ of tree-width at most $k$ has at most
	$k\cdot\left|V(G)\right|$ edges.
	\qed\end{fct}

Fact~\ref{fact:number-of-edges} can be shown by induction on the number of vertices
(see e.g.\ \cite{FG06}).
We will make frequent use of the following 
well-known fact about tree decompositions (see~\cite{die06}): 

\begin{fct}\label{fact:cliques}
Let $(T,B)$ be a tree decomposition of some graph $G$, 
and let $C\subseteq V(G)$. 
If for all $v,w\in C$, some piece of $(T,B)$ covers both $v$ and $w$, 
then there is some piece $B_t$ covering $C$ entirely, i.e.\ $C\subseteq B_t$. 

In particular, every clique in $G$ is covered by some piece. 
\qed\end{fct}


\section{First order tree-width}\label{section:fotw}

\subsection{Stratified tree-width}

We start with defining stratified tree-width. 
Then, first order tree-width is defined as a special case. 
Although it is our only application of stratified tree-width, 
stating results in greater generality allows us to focus on their essence. 
It is also quite possible, that further applications will arise in the future. 

Any rooted tree $T$ induces a natural partial order $<_T$ on its nodes, where
the smallest element is the root. 
For a tree decomposition $(T,B)$ of a graph G and a vertex $v\in V(G)$, 
let $t_v\in T$ denote the $<_T$-minimal tree node that covers $v$. 
By (TD3), the node $t_v$ is well-defined. 
Now, let $d\colon V(G)\to\mathbb N$ be a function. 
We say that a tree decomposition $(T,B)$ of $G$ is $d$-\emph{stratified}, 
if all $u,v\in V(G)$
with $t_u<_T t_v$ satisfy $d(u)\leq d(v)$. 
The \emph{tree-width of $(G,d)$} is defined as
\[
 \tw(G,d):= \min\big\{\w(T,B)\bigmid
   \textsl{$(T,B)$ is a $d$-stratified tree decomposition of $G$}\big\}.
\]
It will sometimes be convenient to work 
with an alternative characterization of stratified tree width: 
let $G=(V,E)$ be a graph and $d:V\to\mathbb N$.
An \emph{elimination ordering} of $(G,d)$ 
is a linear ordering $(v_1,\ldots, v_n)$ of $V$ which \emph{respects} $d$, 
i.e.\ $i<j$ implies $d(v_i)\leq d(v_j)$.
With an elimination ordering we associate a sequence of graphs as follows: 
\begin{iteMize}{$\bullet$}
\item $G_n:=G$
\item $V(G_{i-1}):=V(G_i)\setminus \{v_i\}$, and
\item $E(G_{i-1}):=\big\{e\in E(G_i)\bigmid v_i\not\in e\big\} \cup
  \big\{\{u,w\}\bigmid u\neq w,\; \{u,v_i\},\{v_i,w\}\in E(G_i)\big\}$
  for $1<i\le n$. 
\end{iteMize}
The \emph{width} of the elimination ordering is $\max_{i\in[n]}\{\deg(v_i)\text{ in }G_i\}$.
The \emph{elimination-width} of $(G,d)$, $\ew(G,d)$, is the minimum width of 
an elimination ordering of $(G,d)$. 
It is well-known that the tree-width of a graph $G$ equals the elimination-width of
$G$ (see~\cite{Arnborg85}), and this fact can be generalised to our setting.

\begin{thm}\label{theo:fotw-ew}
Let $G$ be a graph and $d:V(G)\to\mathbb N$.
Then $\tw(G,d)=\ew(G,d).$ 
\end{thm}

\proof
Towards a proof of $\tw(G,d)\geq\ew(G,d)$, let $(T,B)$ be a $d$-stratified
tree decomposition for $G$ of width $k$. We may assume that
$(T,B)$ is \emph{small}, i.e.~all nodes $s,t\in V(T)$ with $s\neq t$ satisfy
$B_s\not\subseteq B_t$. Recall that for a vertex $v\in V(G)$, 
$t_v$ denotes the $<_T$-minimal node of $T$ with $v\in B_t$.
We now define an ordering $v_1,\ldots,v_n$ of $V(G)$ 
such that for all $1\leq i,j\leq n$ we have
\begin{iteMize}{$\bullet$}
   \item $i<j$ implies $d(v_i)\leq d(v_j)$, 
	\item there is a piece $B_i$ of $(T,B)$
		containing $v_i$ and all the neighbours of $v_i$ in $G_i$. 
\end{iteMize}
In particular, $v_1,\ldots,v_n$ is an elimination ordering of $(G,d)$
of width at most~$k$.

\begin{clam}\label{claim:ex x d-max}
There exists a vertex $v\in V(G)$ with $d(v)$ maximum, 
such that $v$ appears in exactly one piece $B_{\ell}$ of $(T,B)$, 
and $\ell$ is a leaf. 
\end{clam}

\proof
Choose any vertex $w$ with $d(w)$ maximum. 
If $w$ is contained in a piece $B_t$ of $(B,T)$ where $t$ is not a leaf 
(otherwise we are done), 
then choose a leaf $\ell\geq_Tt$ of $T$. 
Let $s$ be the parent of $\ell$. 
Choose $v\in B_{\ell}\setminus B_s$ 
(such a $v$ exists since the decomposition is small). 
Since $(T,B)$ is $d$-stratified and $t_w<_T\ell=t_v$, we have 
$d(w)\leq d(v)$, and hence by maximality $d(w)=d(v)$, proving the claim. 
\cqed

Let $v_n:=v$. 
Then we replace $G$ by $G_{n-1}$, 
we restrict $d$ and $(T,B)$ to $G_{n-1}$ and we proceed by induction. 

Towards $\tw(G,d)\leq\ew(G,d)$, let $v_1,\ldots,v_n$ be an ordering of $V(G)$ of width
at most $k$ and let $G_1,\ldots, G_n=G$ be the associated sequence of graphs. 
For $i=1,\ldots,n$ we define tree decompositions for the $G_i$ that respect $d$ and have 
width at most $k$. For $i=1$ we take the trivial decomposition. Given a tree decomposition
of $G_{i-1}$, we choose a piece containing all the neighbours of $v_i$ in $G_{i}$ 
(such a piece exists, because the neighbours induce a clique in $G_{i-1}$), 
and we attach to it a new piece containing $v_i$ and all neighbours of $v_i$ in $G_i$. 
Let $(T,B)$ be the tree decomposition obtained for $G=G_n$. Obviously,
$(T,B)$ has width at most $k$. Moreover, $(T,B)$ is $d$-stratified: let
$v_i,v_j\in V(G)$. If $t_{v_i}<_Tt_{v_j}$, then, by construction, we
have $i<j$. Since $v_1,\ldots,v_n$ is an elimination ordering of $(G,d)$, 
this implies $d(v_i)\leq d(v_j)$. 
\qed

\subsection{First order tree-width}

For a formula $\phi$, the \emph{formula graph} is the undirected graph $\gfi$,
with vertices $\var(\phi)$, and edges $\{x,y\}$ whenever $x$ and $y$ are free variables, or
when $x$ and $y$ occur together in some atom of $\phi$. 
(If $\phi$ is not straight, then we obtain the 
formula graph of $\phi$ by first making it straight.) 
Note that the formula graph depends on the
syntax of the formula. Logically equivalent formulae may have different formula graphs.

We now introduce a partial order $\preceq_{\phi}$ on the variables of a
formula $\phi$, from which we then obtain the \emph{essential alternation depth},
$\ad(x)$, of a variable $x\in\var(\phi)$. 
Given a tree decomposition
of $\gfi$ of width $k-1$ that respects $\ad$, 
we show in Section~\ref{section:mc}, how to transform the formula $\phi$
bottom up along the decomposition into an equivalent $\fok$-formula.
In this transformation, we want to `reuse' as many variables as possible,
so, intuitively, the `worst case' is that $\phi$ is in prenex normal form. 
Hence we want to 
`undo' prenex normal form, pushing quantifiers as far as possible 
away from the root 
in the syntax tree. Of course, we have to make sure that we 
obtain an equivalent formula. Intuitively, $\preceq_{\phi}$
gives us a partial order of quantifications that we
have to respect while undoing prenex normal form.

For a bound variable $x$, let $Q_x\in\{\exists,\forall\}$ 
be the type of quantifier used to quantify $x$ in $\phi$. 
Then the \emph{scope} of $x$ 
is the unique subformula $\psi$ of~$\phi$ 
such that $Q_xx\psi$ is a subformula of $\phi$. 
For bound variables $x,y$ of $\phi$, we write $x\leq_{\phi}y$ to denote 
that $x=y$ or $y$ is quantified in the scope of~$x$. 
For a set $X$ of variables, we use $\phi_{[X]}$ to denote 
the minimal (with respect to subformulaship) subformula of $\phi$ 
which contains all atoms using variables from $X$. 

\begin{defi}
Let $\trianglelefteq$ be a binary relation 
on the variables of some formula $\phi$. 
Then two variables $x$ and $y$ are 
\emph{entangled} with respect to $\trianglelefteq$ and $\phi$, 
if $x$ occurs in $\phi_{[y\trianglelefteq]}$ 
and $y$ occurs in $\phi_{[x\trianglelefteq]}$ 
(as usual, we use $x\trianglelefteq$ 
to denote $\{x'\mid x\trianglelefteq x'\}$). 
\end{defi}

\begin{defi}\label{defi:preceq}
Let $\phi$ be a straight formula. 
Then $\preceq_{\phi}$ is the minimal (with respect to $\subseteq$) 
binary relation on $\var(\phi)$, 
such that the following hold. 
\begin{enumerate}[(1)]
\item $\preceq_{\phi}$ is reflexive. 
\item $\preceq_{\phi}$ is transitive. 
\item If $x\leq_{\phi}y$, $Q_x\not=Q_y$ 
  and there is a sequence $x=z_0,\ldots,z_n=y$ of bound variables 
  such that for all $0\leq i<n$ we have that $z_i,z_{i+1}$ are entangled
  with respect to $\preceq_{\phi}$ and $\phi$ 
  and that $x\preceq_{\phi}z_i$ or $y\preceq_{\phi}z_i$, 
  then $x\preceq_{\phi}y$ (\emph{\alt}). 
\end{enumerate}
\end{defi}

\noindent In order to see that $\preceq_{\phi}$ is well-defined, 
observe that Definition~\ref{defi:preceq} is in fact an inductive definition: 
all three conditions can be restated as closure of $\preceq_{\phi}$ 
under some operator on binary relations, 
and all three operators are monotone with respect to $\subseteq$. 
The least obvious case is the one of the operator underlying \alt. 
To establish monotonicity in this case, 
assume that $\trianglelefteq$ and $\trianglelefteq'$ 
are binary relations on $\var(\phi)$ 
and that $(\trianglelefteq) \subseteq (\trianglelefteq')$. 
We have to show that whenever two variables $x,y$ 
satisfy \alt{} with respect to $\trianglelefteq$, 
then they also do with respect to $\trianglelefteq'$. 
For any variable $z$ 
we have $(z\trianglelefteq) \subseteq (z\trianglelefteq')$, 
so $\phi_{[z\trianglelefteq]}$ is a subformula of $\phi_{[z\trianglelefteq']}$. 
Thus, entanglement of some variables 
with respect to $\trianglelefteq$ and $\phi$ 
implies entanglement with respect to $\trianglelefteq'$ and $\phi$. 
Hence any witness for \alt{} with respect to $\trianglelefteq$ 
is also one with respect to $\trianglelefteq'$. 

At many places, we will use proof by induction 
on the inductive definition of $\preceq_{\phi}$. 
Therefore, we explicate how the inductive principle works in this case. 

\begin{lem}\label{lem:induction}
Let $\phi$ be a formula and $P$ a property of pairs of variables from $\phi$. 
If 
\begin{enumerate}[\em(1)]
\item $P(x,x)$ holds for all $x\in\var(\phi)$, 
\item $x\preceq_{\phi}y$, $y\preceq_{\phi}z$, $P(x,y)$ and $P(y,z)$ 
  imply $P(x,z)$, and 
\item if $x\leq_{\phi}y$, $Q_x\not=Q_y$, 
  $(\trianglelefteq)\subseteq(\preceq_{\phi})$ 
  such that $P(x',y')$ holds for all $x'\trianglelefteq y'$, 
  and for some sequence $x=z_0,\ldots,z_n=y$ and all $0\leq i<n$ 
  we have $z_i\in(x\trianglelefteq)\cup(y\trianglelefteq)$ 
  and entanglement of $z_i$ and $z_{i+1}$ 
  with respect to $\trianglelefteq$ and $\phi$, 
  then $P(x,y)$, 
\end{enumerate}
then $P(x,y)$ holds for all $x,y$ such that $x\preceq_{\phi}y$. 
\qed\end{lem}

\begin{rem}\label{rem:preceq-properties}\hfill
\begin{enumerate}[(1)]
\item The relation $\preceq_{\phi}$ is a subrelation of $\leq_{\phi}$: 
  $(\preceq_{\phi})\subseteq(\leq_{\phi})$,
\item the relation $\preceq_{\phi}$ is a partial order, and
\item $x\preceq_{\phi}y$ holds whenever $x$ and $y$ are entangled, 
  $Q_x\not=Q_y$, and $x\leq_{\phi}y$.
\end{enumerate}
\end{rem}

\proof
1 follows since $\leq_{\phi}$ satisfies all closure conditions. \\
2: $x\preceq_{\phi}y$ is reflexive and transitive by definition, 
and it inherits anti-symmetry from $(\leq_{\phi})$ by~1. \\
3: this follows by letting $n=1$ in \alt{}.
\qed

\begin{exa}\label{ex:preceq}
Let $\phi:=\exists x\forall y \exists z \big(Pxy\wedge\forall u(Ryu\vee Pzu)\big)$.
Then $x\preceq_{\phi}y$ and $z\preceq_{\phi}u$
by Remark~\ref{rem:preceq-properties},~3, 
$y\preceq_{\phi}z$ by Alternation (witnessed by the sequence $y,u,z$), 
and $x\preceq_{\phi}z$, $y\preceq_{\phi}u$, and $x\preceq_{\phi}u$ 
by Transitivity.
In this example, all entanglements are due to 
the two variables in question occuring in the same atom. 
\end{exa}

We use $\phi_x$ as a shorthand for $\phi_{[x\preceq_{\phi}]}$, 
and we say that $x$ and $y$ are entangled in $\phi$, 
if they are entangled with respect to $\preceq_{\phi}$ and $\phi$. 
Note that this is the case 
if and only if $x$ occurs in $\phi_y$ and $y$ occurs in $\phi_x$. 
Observe further that $\phi_x$ is a subformula of the scope of $x$. 
The idea behind entanglement is to capture interaction between variables. 

\begin{exa}\label{exa:entanglement}
Let $\phi:=\forall x\forall x'\exists y
  (((Py\wedge Px)\vee Px) \wedge ((Py\wedge Px')\vee Px'))$. 
Then $\phi_{[\{y\}]}$ already is the whole quantifier free part of $\phi$, 
hence so is $\phi_y$. 
Further, $\phi_x$ contains $\phi_{[\{x\}]}=(Py\wedge Px)\vee Px$. 
Thus $x$ occurs in $\phi_y$ and $y$ occurs in $\phi_x$, 
so $x$ and $y$ are entangled. 
It follows that $x\preceq_{\phi}y$, so $\phi_x$ contains $\phi_y$. 
As this is the whole quantifier free part, we have $\phi_x=\phi_y$. 
In a similar way we obtain $\phi_{x'}=\phi_y$. 
Consequently, $x$ and $x'$ are entangled as well. 
Intuitively, $x$ and $x'$ interact through $y$. 
\end{exa}

\begin{exa}\label{exa:non-entanglement}
Let $\phi:=\forall x\exists y\forall z(Rzy\vee(Px\wedge Py))$. 
Then $y$ and $z$ are entangled, because they occur in the same atom. 
$x$, however, is not entangled with any other variable, 
because $\phi_{[\{x\}]}=Px$ does not contain any variable besides $x$. 
The same holds for $\psi:=\exists y(\forall zRzy\vee(\forall xPx\wedge Py))$, 
which illustrates that $x$ does not interact at all. 
Thus, $(\preceq_{\phi})=\{(y,z),(x,x),(y,y),(z,z)\}$. 
\end{exa}

\begin{exa}\label{exa:ent.: dep. on reordering}
For $n>0$ let 
\[\phi_n:=\exists x_1\ldots\exists x_n\forall y\exists z
  \left(\bigwedge\limits_{1\leq i\leq n}Rx_iz\wedge Py\right)\] 
and 
\[\psi_n:=\exists x_1\ldots\exists x_n\forall y\exists z
  \left(\bigwedge\limits_{1\leq i\leq n}(Rx_iz\wedge Py)\right)\,.\] 
In $\phi_n$, the only entanglements are between the $x_i$ and $z$. 
Consequently, $\preceq_{\phi_n}$ is the equality relation on $\var(\phi_n)$. 
On the other hand, both $(\psi_n)_y$ and $(\psi_n)_z$ 
coincide with the quantifier free part of $\psi_n$, 
so $y$ and $z$ are entangled in $\psi_n$. 
It follows that $y\preceq_{\psi_n}z$. 
Nevertheless, as $y$ does not occur in $(\psi_n)_{x_i}=Rx_iz$, 
$y$ is not entangled with $x_i$ in $\psi_n$. 
\end{exa}

\begin{defi}\label{defi:ead} 
	Let $\phi$ be a first order formula and $x\in\var(\phi)$. 
	The \emph{essential alternation depth} of $x$ in $\phi$, denoted by $\ad(x)$,
	is the maximum over all $\preceq_{\phi}$-paths $P$ ending in $x$ 
	of the number of quantifier changes in $P$, 
	adding $+1$ in case the first variable on $P$ is existentially quantified 
	and $+2$ if it is universally quantified. 
	If $x$ is a free variable, we let $\ad(x)=0$.
\end{defi}

The $+1$ respectively $+2$ in the definition 
makes sure that $\ad(x)$ is odd if and only if $Q_x=\exists$. 

\begin{exa}
The formula from Example~\ref{ex:preceq} 
satisfies $\ad(x)=1$, $\ad(y)=2$, $\ad(z)=3$, and $\ad(u)=4$. 

The formula from Example~\ref{exa:entanglement} 
satisfies $\ad(x)=\ad(x')=2$ and $\ad(y)=3$. 

The formulae from Example~\ref{exa:non-entanglement} 
satisfy $\ad(x)=\ad(z)=2$, $\ad(y)=1$, and $\dd_{\psi}=\ad$. 

For the formulae from Example~\ref{exa:ent.: dep. on reordering} 
we have $\dd_{\phi_n}(x_i)=\dd_{\psi_n}(x_i)=\dd_{\phi_n}(z)=1$, 
$\dd_{\phi_n}(y)=\dd_{\psi_n}(y)=2$, and $\dd_{\psi_n}(z)=3$. 
\end{exa}

\begin{defi}\label{defi:ad} 
If we replace $\preceq_{\phi}$ by $\leq_{\phi}$ in Definition~\ref{defi:ead}, 
we obtain the (usual) \emph{alternation depth} of $x$ in $\phi$, 
which we denote by $\altfi$.
\end{defi}

\begin{rem}\label{rem:ead-ad-sigmat}
	Every formula $\phi$ satisfies $\ad\leq\altfi$. 
\end{rem}

\begin{exa}
For the formuale from Examples \ref{ex:preceq}~and~\ref{exa:entanglement} 
we have $\altfi=\ad$. 

The formulae from Example~\ref{exa:non-entanglement} 
satisfy $\altfi(x)=2$, $\altfi(y)=3$, $\altfi(z)=4$, 
and $\alte_{\psi}=\dd_{\psi}$. 

For the formulae from Example~\ref{exa:ent.: dep. on reordering} 
we have $\alte_{\phi_n}=\alte_{\psi_n}=\dd_{\psi_n}$. 
\end{exa}

\begin{defi}[First order tree-width]\label{defi:fotw} 
	For a formula $\phi$ we define the \emph{first order tree-width}
	of $\phi$ by $\fotw(\phi):=\tw(\gfi,\ad)$.\\ 
Accordingly, we say that $(T,B)$ is a \emph{tree decomposition for $\phi$}, 
if $(T,B)$ is an $\ad$-stratified tree decomposition for $\gfi$. 
\end{defi}

Note that for a formula $\phi$, the
variables $\free(\phi)$, as well as the variables of any atom or literal in
$\phi$ induce cliques in $\gfi$. 
In particular, since $\ad(x)=0$ for any free
variable $x$, by Fact~\ref{fact:cliques}
it is no restriction to require that the free variables be
covered in the root of a tree decomposition. 

In general, the difference between $\fotw(\phi)$ and $\tw(\gfi)$ can be
unbounded: 

\begin{prop}\label{prop:unbounded}
For every $n>0$ there is a
formula $\phi_n$ with $\fotw(\phi_n)=n$ and $\tw(G_{\phi_n})=1$. 
\end{prop}

\proof 
Let
\[
  \phi_n = \exists x_1 \ldots \exists x_n \forall y
    \bigwedge\limits_{1\leq i\leq n}Ex_iy \,.
\]
Then $G_{\phi_n}$ is the $n$-star with center $y$, 
and we have $\dd_{\phi_n}(x_i)=1$ for all $1\leq i\leq n$, 
and $\dd_{\phi_n}(y)=2$. 
It is easy to see 
that any $\dd_{\phi_n}$-stratified tree decomposition $(T,B)$ of $G_{\phi_n}$ 
has a piece $\{x_1,x_2,\ldots,x_n,y\}$, namely $B_{t_y}$. 
On the other hand, such a tree decomposition needs no other pieces. 
Hence $\fotw(\phi_n)=n$. 
Since $G_{\phi_n}$ is a tree we have $\tw(G_{\phi_n})=1$.
\qed

\begin{lem}\label{lem:ead poly}
Given a formula $\phi$, 
we can compute $\preceq_{\phi}$ and $\ad$ in polynomial time. 
\end{lem}

\proof
In order to compute $\preceq_{\phi}$, 
consider the three closure operators implicit in its definition. 
As $\preceq_{\phi}$ is a binary relation on $\var(\phi)$, 
a quadratic number of applications of the closure operators 
suffices to produce $\preceq_{\phi}$. 
Hence it remains to show that each closure operator 
is computable in polynomial time. 
This is immediate for \refl{} and \trans. 
For \alt, let $(\trianglelefteq)\subseteq\var(\phi)^2$ 
be the current approximation of $\preceq_{\phi}$. 
First, we compute the formulae $\phi_{[z\trianglelefteq]}$ 
for all variables $z$, 
and from these the entanglement relation. 
Then, checking whether some pair $(x,y)$ 
needs to be added to $R$ because of \alt{} 
basically amounts to reachability in the entanglement graph 
restricted to $(x\trianglelefteq)\cup(y\trianglelefteq)$. 

It is clear that $\ad$ can be computed 
from $\phi$ and $\preceq_{\phi}$ in polynomial time. 
\qed

\subsection{Xenerp normal form}

Prenex normal form aims 
to make the scopes of quantifiers as large as possible. 
Working in the opposite direction, we obtain what we term xenerp normal form. 

\begin{defi}
A subformula $\chi$ of a formula $\phi$ 
is in \emph{xenerp normal form with respect to $\phi$}, 
if for all variables $x$ quantified in $\chi$ the following holds: 
$\phi_x$ is immediately preceeded 
by a quantifier sequence which contains $Q_xx$. 

A formula $\phi$ is in \emph{xenerp normal form}, 
if it is in xenerp normal form with respect to itself. 
\end{defi}

\begin{exa}\label{ex:xenerp}
Recall the formulae $\phi$ and $\psi$ from Example~\ref{exa:non-entanglement}. 
We have already seen that $\ad=\dd_{\psi}$. 
Furthermore, $\phi$ and $\psi$ are equivalent 
and $\psi$ is in xenerp normal form, 
whereas $\phi$ is not in xenerp normal form. 
\end{exa}

The following lemma presents equivalent transformations of formulae, 
such that neither the formula graph, nor the essential alternation depth, 
nor the corresponding tree decompositions change. 

\begin{lem}\label{lem:replacements}
Let $\phi$ and $\psi$ be formulae satisfying either 1,~2 or~3. 
\begin{enumerate}[\em(1)]
\item There are formulae $\chi_1,\ldots,\chi_n$, 
  a positive Boolean combination $\theta$ of $n$ arguments, 
  and a variable $x$ which does not occur in $\chi_2,\ldots,\chi_n$, 
  such that $\psi$ is obtained from $\phi$ by replacing 
  a subformula $\theta(Q_x x\chi_1,\chi_2,\ldots,\chi_n)$ 
  by $Q_x x\theta(\chi_1,\chi_2,\ldots,\chi_n)$. 

\item There are a formula $\chi$, and variables $x,y$ with $Q_x=Q_y$ 
  such that $\psi$ is obtained from $\phi$ by replacing 
  a subformula $Q_xxQ_yy\chi$ by $Q_yyQ_xx\chi$. 

\item There are a formula $\chi$, 
  and variables $x,y$ with $\phi_x$ a proper subformula of $\phi_y$, 
  such that $\psi$ is obtained from $\phi$ by replacing 
  a subformula $Q_xxQ_yy\chi$ by $Q_yyQ_xx\chi$, 
  where $Q_yy\chi$ is xenerp with respect to $\phi$. 
\end{enumerate}
Then $\phi\equiv\psi$, $\ad=\dd_{\psi}$, and $\gfi=G_{\psi}$. 
Consequently, 
tree decompositions for $\phi$ coincide with tree decompositions for $\psi$ 
and in particular $\fotw(\phi)=\fotw(\psi)$. 

Implicitly, we assume in these cases that $Q_x$ and $Q_y$ 
are the same with respect to the formula $\phi$ 
and with respect to the formula $\psi$. 
\end{lem}

\proof
\setcounter{clam}0
In all three cases the formulae differ only in the position of quantifiers, 
so $\gfi=G_{\psi}$ is immediate. 
Also, for all $X\subseteq\var(\phi)=\var(\psi)$, 
the formulae $\phi_{[X]}$ and $\psi_{[X]}$ are essentially equal: 
the only potential difference between them 
is the same shift of quantifiers which led from $\phi$ to $\psi$. 
In particular, the same variables occur in $\phi_{[X]}$ as in $\psi_{[X]}$. 
Hence, all differences between $\preceq_{\phi}$ and $\preceq_{\psi}$ 
(and thus between $\ad$ and $\dd_{\psi}$) 
must ultimately stem from differences between $\leq_{\phi}$ and $\leq_{\psi}$. 

For the first replacement, the equivalence $\phi\equiv\psi$ is well-known. 
For the other parts of the statement, 
the only change between $\leq_{\phi}$ and $\leq_{\psi}$ is, 
that for all variables $y$ quantified in $\chi_i$ for some $2\leq i\leq n$, 
we have $x\not\leq_{\phi}y$ but $x\leq_{\psi}y$. 
We show that this change has no impact on $\preceq$. 
Clearly $(\preceq_{\phi})\subseteq(\preceq_{\psi})$ 
because of $(\leq_{\phi})\subseteq(\leq_{\psi})$. 
For the converse we use induction on derivations. 
More precisely, 
we show by induction on the inductive definition of $\preceq_{\psi}$ 
that for all $x'\preceq_{\psi}y'$ we have $x'\preceq_{\phi}y'$. 
That is the inductive property $P(x',y')$ as in Lemma~\ref{lem:induction} 
is $x'\preceq_{\phi}y'$. 
The first two inductive rules are trivial, 
because we know that $\preceq_{\phi}$ is reflexive and transitive. 
Hence we can concentrate on \alt{}. 
So let $x'\leq_{\psi}y'$, 
$(\trianglelefteq)\subseteq(\preceq_{\psi})\cap(\preceq_{\phi})$, 
and $x'=z_0,\ldots,z_n=y'$ be given such that $Q_x\not=Q_y$, 
and for all $0\leq i<n$ 
we have $z_i\in(x'\trianglelefteq)\cap(y'\trianglelefteq)$ 
and entanglement of $z_i,z_{i+1}$ with respect to $\trianglelefteq$ and $\psi$. 
By our above observation on the similarity of $\phi_{[X]}$ and $\psi_{[X]}$ 
for any set $X$ of bound variables, 
it follows that $z_i$ and $z_{i+1}$ 
are also entangled with respect to $\trianglelefteq$ and $\phi$, 
and thus with respect to $\preceq_{\phi}$ and $\phi$. 
Consequently, either $x'\preceq_{\phi}y'$ holds (and we are done), 
or $x'\not\leq_{\phi}y'$. 
Then, together with $x'\leq_{\psi}y'$ it follows that $x'=x$ 
and $y'$ is quantified in $\chi_j$ for some $2\leq j\leq n$. 
As $x\leq_{\phi}z_i$ or $y'\leq_{\phi}z_i$ for all $0\leq i\leq n$, 
all $z_i$ are quantified either in $Q_xx\chi_1$ or in $\chi_j$. 
As the former is true for $z_0=x$ and the latter for $z_n=y'$, 
there is some $0\leq i<n$ 
such that $z_i$ is quantified in $Q_xx\chi_1$ and $z_{i+1}$ in $\chi_j$. 
Then $\phi_{[z_i\trianglelefteq]}$ is a subformula of $\phi_{z_i}$, 
which in turn is a subformula of the scope of $z_i$ and thus of $Q_xx\chi_1$. 
Similarly, the scope of $z_{i+1}$ is a subformula of $\chi_j$. 
As all occurences of $z_{i+1}$ are in its scope 
and $\chi_j$ is disjoint from $Q_xx\chi_1$, 
this contradicts the fact that $z_{i+1}$ occurs in $\phi_{[z_i\trianglelefteq]}$.

In the second replacement, the equivalence $\phi\equiv\psi$ also is well-known. 
So let us show $(\preceq_{\phi})=(\preceq_{\psi})$. 
As the replacement is symmetric, 
it suffices to show $(\preceq_{\psi})\subseteq(\preceq_{\phi})$, 
which we do by induction. 
Again, the cases of \refl{} and \trans{} are clear. 
For \alt, let $\trianglelefteq$ and $x'=z_0,\ldots,z_n=y'$ be given as above. 
Again, we obtain that \alt{} also yields $x'\preceq_{\phi}y'$ 
unless $x'\not\leq_{\phi}y'$. 
But then $x'=x$ and $y'=y$ contradicting $Q_x=Q_y$.

In the third replacement, 
if $Q_x=Q_y$ then this case is subsumed by the second replacement 
so we may assume $Q_x\not=Q_y$. 
Let us start by showing $\phi\equiv\psi$. 
We will even show $Q_xxQ_yy\chi\equiv Q_yyQ_xx\chi$. 
As $Q_yy\chi$ is in xenerp normal form we have, 
for variables $v,w$ quantified in $Q_yy\chi$,
that $\phi_w$ is a subformula of $\phi_v$ whenever $v\leq_{\phi}w$. 
  In this case $w$ occurs in $\phi_v$ 
  (as the formula is straight, $w$ does occur somewhere, 
  and it can only occur in $\phi_w$ which is a subformula of $\phi_v$). 
So for entanglement of $v$ and $w$ 
it suffices to show that $v$ occurs in $\phi_w$. 
Let $V:=\{v\in \var(\phi)\mid  y\leq_{\phi}v\text{ and }\phi_x\text{ is a subformula of }
\phi_v\}$. 
Observe, that $V$ is linearly ordered by $\leq_{\phi}$.

\begin{clam}
There is a subformula $\theta$ of $\chi$ 
which is a superformula of $\phi_x$ 
such that no variable from $V$ occurs free in $\theta$. 
\end{clam}

\proof
For contradiction, assume the opposite. 
Then we inductively define a sequence $v_0,v_1,\ldots$ of variables. 
Take $v_0$ to be $x$. For all $i\geq 1$ we will have $v_i\in V$. 
For defining $v_{i+1}$ from $v_i$, 
let $\theta_i$ be $\phi_{v_i}$, 
together with the sequence of quantifications of variables from $V$ 
which immediately preceeds $\phi_{v_i}$. 
Hence $\theta_i$ is a subformula of $Q_yy\chi$. 
If $\theta_i=Q_yy\chi$, then we terminate the sequence. 
Otherwise, by our assumption, 
some variable from $V$ occurs free in $\theta_i$. 
Then let $v_{i+1}$ be such a variable. 
As $Q_yy\chi$ is xenerp, 
$\theta_i$ is a proper subformula of $\phi_{v_{i+1}}$. 
This implies that $v_i$ occurs in $\phi_{v_{i+1}}$ 
and the converse holds by choice of $v_{i+1}$, 
because $\theta_i$ essentially coincides with $\phi_{v_i}$. 
Hence $v_i$ and $v_{i+1}$ are entangled whenever both are defined. 
If furthermore $i\not=0$, 
then we also have $v_{i+1}\leq_{\phi} v_i$ and $v_{i+1}\not=v_i$, 
because $Q_yy\chi$ is xenerp 
and $\phi_{v_i}$ is a proper subformula of $\phi_{v_{i+1}}$. 
Hence, as $V$ is finite, the sequence $v_0,\ldots$ must terminate, 
say with $v_n$. 
As $\phi_x$ is a proper subformula of $\phi_y$, 
we have $\theta_0\not=Q_yy\chi$, so $n>0$. 
Now let $0\leq m\leq n$ be minimal such that $Q_{v_m}=Q_y$. 
The case that there are no such $m$ will be handled later. 
As $Q_{v_0}=Q_x\not=Q_y$, we have $m>0$ 
so $\phi_x$ is a proper subformula of $\phi_{v_m}$. 
We obtain $v_m\preceq_{\phi}v_i$ for all $1\leq i\leq m$ 
using a backwards induction as follows. 
The base case is \refl. 
For the inductive step we have a chain $v_m,\ldots,v_i$ of entanglements 
and by the inductive hypothesis we have $v_m\preceq_{\phi}v_j$ 
for all intermediate $j>i$. 
Further $v_m\leq v_i$ and $Q_{v_m}=Q_y\not=Q_{v_i}$, 
so we obtain $v_m\preceq_{\phi}v_i$ using \alt. 
Then, in total we have a chain $x=v_0,v_1,\ldots,v_m$ of entanglements 
such that $x\preceq_{\phi}v_0$ by \refl, 
$v_m\preceq_{\phi}v_i$ for all $1\leq i<m$ 
and $Q_x\not=Q_y=Q_{v_m}$, so \alt{} implies $x\preceq_{\phi}v_m$. 
But then $(v_m\preceq_{\phi})\subseteq(x\preceq_{\phi})$, 
contradicting that $\phi_x$ is a proper subformula of $\phi_{v_m}$. 
Now for the case that $Q_{v_i}=Q_x$ for all $i$ such that $v_i$ is defined. 
Then $\theta_n=Q_yy\chi$, so $\phi_{v_n}=\phi_y$ 
and thus $v_n$ and $y$ are entangled. 
This time we have a chain $y,v_n,\ldots,v_1,x$ of entanglements 
with the same properties as the sequence $v_m,\ldots,v_1,x$ above: 
it is descending with respect to $\leq_{\phi}$ except for the last step, 
the first variable has the same quantifier as $y$ 
and all other variables have the same quantifier as $x$. 
Thus similar to the above we obtain $x\preceq_{\phi}y$, 
this time in contradiction to $\phi_x$ being a proper subformula of $\phi_y$. 
This proves the claim.
\cqed

So let $\theta$ be a subformula of $\chi$ and a superformula of $\phi_x$ 
such that no variable from $V$ occurs free in $\theta$. 
Hence all free variables of $\theta$ except $x$ 
are also free variables of $Q_xxQ_yy\chi$. 
Without loss of generality we may assume that $Q_x=\exists$ and $Q_y=\forall$. 
It is well-known that $\exists x\forall y\chi$ 
implies $\forall y\exists x\chi$. 
For the converse, 
let $\xI$ be an interpretation for $\exists x\forall y\chi$ 
such that $\xI\models\forall y\exists x\chi$. 
We need to show that $\xI\models\exists x\forall y\chi$. 
For a value $a$ from the universe of the interpretation $\xI$, 
we have that $\xI\frac ax$ is an interpretation for $\forall y\chi$.
As all free variables of $\theta$ are also free in $\forall y\chi$, 
we have that $\xI\frac ax$ also is an interpretation for $\theta$. 
Now let $a_0$ be some value such that $\xI\frac{a_0}x\models\theta$. 
If no such $a_0$ exists, let instead $a_0$ be arbitrary. 
Now for all $a$ we have that $\xI\frac ax\models\theta$ 
implies $\xI\frac{a_0}x\models\theta$. 
Now let $b$ be an arbitrary value. 
As $\xI\models\forall y\exists x\chi$, we have that 
$\xI_1:=\xI\frac by\frac{a'}x\models\chi$ for some value $a'$. 
Let us examine the impact that replacing $\xI_1$ 
by $\xI_2:=\xI\frac by\frac{a_0}x$ has on the formula $\chi$. 
Recall that $\phi_x$ is a subformula of $\theta$, 
so in particular $x$ does not occur free in $\chi$ except in $\theta$. 
Hence the only possible change comes from $\theta$. 
If $\xI_1\models\theta$, then also $\xI_2\models\theta$ and nothing changes, 
that is $\xI_2\models\chi$. 
The same argument holds if $\xI_1,\xI_2\not\models\theta$. 
Otherwise $\xI_1\not\models\theta$ and $\xI_2\models\theta$. 
In this case recall, that all formulae are in negation normal form, 
so $\chi$ is positive in $\theta$. 
Thus $\xI_1\models\chi$ again implies $\xI_2\models\chi$. 
So $\xI\frac{a_0}x\frac by=\xI\frac by\frac{a_0}x\models\chi$ 
for all values $b$, 
which implies $\xI\frac{a_0}x\models\forall y\chi$ 
and then $\xI\models\exists x\forall y\chi$. 

Next, let us compare $\preceq_{\phi}$ with $\preceq_{\psi}$. 
The only difference between $\leq_{\phi}$ and $\leq_{\psi}$ is, 
that $x\leq_{\phi}y\not\leq_{\phi}x$ while $x\not\leq_{\psi}y\leq_{\psi}x$. 
To show $(\preceq_{\phi})=(\preceq_{\psi})$, 
we show inclusion in both directions by induction. 
Let us start with $(\preceq_{\phi})\subseteq(\preceq_{\psi})$. 
As in the proofs for the other replacements, 
the only interesting case is 
where $x'\preceq_{\phi}y'$ but $x'\not\leq_{\psi}y'$. 
This implies $x'=x$ and $y'=y$. 
Then $x\preceq_{\phi}y$ 
in contradiction to $\phi_x$ being a proper subformula of $\phi_y$. 
For the other inclusion we need to recall a little more 
from the above cases: 
we obtain a sequence of entanglements 
(which are such both in $\psi$ and in $\phi$) 
$x'=z_0,\ldots,z_n=y'$ such that for all $0\leq i<n$ 
we have $x'\preceq_{\psi}z_i$ and $x'\preceq_{\phi}z_i$ 
or $y'\preceq_{\psi}z_i$ and $y'\preceq_{\phi}z_i$. 
Also we have $x'\leq_{\psi}y'$ and we are done unless $x'\not\leq_{\phi}y'$, 
hence $x'=y$ and $y'=x$. 
But then we have $x\leq_{\phi}y$ and a chain 
$x=v_n,\ldots,v_0=y$ of entanglements. 
As further $x\preceq_{\phi}v_n$ by \refl, 
we obtain $x\preceq_{\phi}y$ which again gives a contradiction.
This concludes the proof of Lemma~\ref{lem:replacements}\qed

Observe, that the first class of replacements in the previous lemma 
are exactly what is used in turning a formula into prenex normal form. 
For xenerp normal form, we need the first and third class. 

\begin{cor}\label{lem:prenex}
Let $\phi$ be a formula 
and let $\psi$ be a prenex normal form (obtained in the usual way) of $\phi$. 
Then $\fotw(\phi)=\fotw(\psi)$. 
\qed\end{cor}

\begin{lem}\label{lem:xenerp}
	{}From a formula $\phi$ we can compute in polynomial time 
a formula $\psi$ in xenerp normal form, 
such that $\phi\equiv\psi$, $\gfi=G_{\psi}$, $\ad=\dd_{\psi}$, 
and consequently, tree decompositions for $\phi$ 
coincide with those for $\psi$. 
\end{lem}

\proof
We work by applying replacements from Lemma~\ref{lem:replacements} 
in one direction or the other. 
More precisely, 
we apply replacements of the first kind backwards whenever possible. 
When\-ever no such replacement is applicable, 
then each quantifier sequence ends with some quantifier $Q_xx$ 
such that the scope of $x$ is an atom, a negated atom, 
or a conjunction or disjunction of two subformulae, 
both of which contain $x$. 
In all three cases, $\phi_x$ contains the scope of $x$. 
As the reversed inclusion always holds, 
we have that $Q_xx$ immediately precedes $\phi_x$. 

If the formula $\phi'$ at hand is xenerp, we are done. 
Otherwise let $x$ be $\leq_{\phi'}$-maximal 
such that the quantifier sequence containing $Q_xx$ does not precede $\phi_x$. 
By the above, this sequence does not end with $Q_xx$, 
so the scope of $x$ has the form $Q_yy\chi$. 
By maximality of $x$, $Q_yy\chi$ is xenerp with respect to $\phi'$. 
In particular, 
the quantifier sequence containing $Q_xx$ is followed by $\phi_y$. 
By the choice of $x$ we conclude that $\phi_x\not=\phi_y$, 
so $\phi_x$ is a proper subformula of $\phi_y$. 
Hence a replacement of the third kind is applicable 
(in the forward direction). 

It remains to show, that repeatedly applying these replacements terminates. 
As a first semi-invariant, consider the sum, taken over all variables $x$, 
of the distance that $Q_xx$ has from the root 
in the syntax tree of the formula. 
(Backwards) replacements of the first kind increase this semi-invariant 
while replacements of the third kind do not change it at all. 
On the other hand, replacements of the third kind 
decrease the number of variable pairs $(x,y)$, 
such that $x\leq_{\phi'}y$ and $\phi_x$ is a proper subformula of $\phi_y$. 
As both semi-invariants are polynomially bounded in $|\phi|$, 
and each replacement (including the test for applicability) 
requires only polynomial time (recall Lemma~\ref{lem:ead poly}), 
the procedure runs in polynomial time. 
\qed

\subsection{Comparing ead with ad}\label{section:comparing}

Let $\phi$ be a first order formula. 
As $\preceq_{\phi}$ is coarser than $\leq_{\phi}$, 
it is immediate that $\ad\leq\altfi$. 
However, this does not directly imply 
that $\fotw(\phi)\leq\tw(\gfi,\altfi)$ also holds. 
Variables which are incomparable by $\leq_{\phi}$ 
(and thus by $\preceq_{\phi}$) 
are given an order by $\ad$ and by $\altfi$, 
but not neccessarily the same one. 
Consequently, not every $\altfi$-stratified tree decomposition 
is also $\ad$-stratified. 
Nevertheless $\fotw(\phi)\leq\tw(\gfi,\altfi)$ does hold, 
and it is the purpose of this subsection to show this fact. 

At the core of our proof there is a double induction 
which we cast into two auxiliary lemmata. 
An \emph{entanglement chain} (in some formula $\phi$) 
is, of course, a sequence $v_0,\ldots,v_n$ of variables 
such that for all $0\leq i<n$ the variables $v_i$ and $v_{i+1}$ 
are entangled in $\phi$. 
The entanglement chain is \emph{hanging}, 
if $v_0,v_n\leq_{\phi}v_i$ for all $0<i<n$. 
It is \emph{crossing}, if $Q_{v_0}\not=Q_{v_n}$. 
It is \emph{nice}, 
if $\min_{\leq_{\phi}}(v_0,v_n)\preceq_{\phi}v_i$ for all $0\leq i\leq n$.
(We will only talk about nice chains in contexts 
where the existence of $\min_{\leq_{\phi}}(v_0,v_n)$ is guaranteed a priori.
As a side note it is not hard to see 
that this minimum exists for all hanging chains.)

\begin{lem}\label{lem:bloedes zeug 1}
Let $\phi$ be a formula such that $\leq_{\phi}$ is a total order 
(for example $\phi$ is in prenex normal form). 
Let $x=v_{-m},\ldots,v_{-1},y=v_0,v_1,\ldots,v_n=z$ 
be an entanglement chain without repetitions 
such that the following hold: 
\begin{enumerate}[\em(1)]
\item $Q_x\not=Q_y=Q_z$. 
\item $x\leq_{\phi}v_i$ for all $-m\leq i<n$. 
\item The subchain $y=v_0,v_1,\ldots,v_n=z$ is hanging. 
\item $x\preceq_{\phi}v_i$ for all $-m\leq i\leq 0$, 
  i.e.~the subchain $x=v_{-m},\ldots,v_{-1},v_0=y$ is nice. 
\item For all $0\leq k<\ell\leq n$, 
  if the subchain $v_k,\ldots,v_{\ell}$ is hanging and crossing, 
  then it is also nice. 
\end{enumerate}
Then the chain is nice. 
\end{lem}

\proof
Otherwise assume a counterexample with $n$ minimal. 
Obviously, $n>0$. 
If $n=1$, then we can use the chain 
to derive either $x\preceq_{\phi}z$ or $z\preceq_{\phi}x$ by \alt. 
In the first case we are done, 
in the second case we use \trans{} with $x$ 
to derive $z\preceq_{\phi}v_i$ for all $-m\leq i\leq 0$. 

Hence in the following we may assume $n>1$. 
Let $0<k<n$ be such, that $v_k\leq_{\phi}v_i$ for all $0<i<n$, 
that is the subchains $v_0,\ldots,v_k$ and $v_k,\ldots,v_n$ are hanging. 
If $Q_{v_k}=Q_x$, then the last premise 
gives $y\preceq_{\phi}v_i$ for all $0\leq i\leq k$ 
and $z\preceq_{\phi}v_i$ for all $k\leq i\leq n$. 
The former, together with $x\preceq_{\phi}v_0=y$, 
gives $x\preceq_{\phi}v_i$ for all $0\leq i\leq k$. 
Hence, for all $-m\leq i\leq n$ 
we have $x\preceq_{\phi}v_i$ or $z\preceq_{\phi}v_i$. 
Thus the chain again witnesses $x\preceq_{\phi}z$ or $z\preceq_{\phi}x$, 
so transitivity also gives $x\preceq_{\phi}v_i$ for all $k\leq i\leq n$ 
or $z\preceq_{\phi}v_i$ for all $-m\leq i\leq k$. 

So we are left with the case $Q_{v_k}=Q_y=Q_z$. 
Then the subchain $x=v_{-m},\ldots,v_k$ 
satisfies all conditions of this lemma. 
Minimality of $n$ implies that $x\preceq_{\phi}v_i$ for all $-m\leq i\leq k$. 
Now we set $m'=m+k$, $n'=n-k$, and $v'_i=v_{i+k}$ for all $-m'\leq i\leq n'$. 
The thus shifted entanglement chain again satisfies all conditions, 
so minimality of the counterexample implies 
that the shifted chain is nice. 
Then so is the original chain. 
\qed

\begin{lem}\label{lem:bloedes zeug 2}
Let $\phi$ be a formula such that $\leq_{\phi}$ is a total order. 
Then every hanging and crossing entanglement chain is nice. 
\end{lem}

\proof
Otherwise assume a counterexample $v_0,\ldots,v_n$ with minimal $n$. 
The fact that the chain is crossing prohibits $n=0$. 
If $n=1$, the claim follows from \alt. 
Hence we assume $n>1$. 
Let $0<k<n$ be such, that $v_k\leq_{\phi}v_i$ for all $0<i<n$. 
We have $Q_{v_k}=Q_x$ or $Q_{v_k}=Q_y$, without loss of generality the latter. 
Then by minimality of the counterexample 
the shifted entanglement chain with $v_k$ in the middle 
satisfies all conditions of Lemma~\ref{lem:bloedes zeug 1}. 
Hence the claim follows from that lemma. 
\qed

\begin{thm}\label{thm:fotw und ad}
For all $\phi\in\ml$ we have $\fotw(\phi)\leq\tw(\gfi,\altfi)$. 
\end{thm}

\proof
Without loss of generality, 
we may assume that $\phi$ is in prenex normal form: 
otherwise let $\phi'$ be a prenex normal form of $\phi$ 
such that $\altfi=\alte_{\phi'}$. 
Such a $\phi'$ can be obtained by moving quantifiers to the left in $\phi$ 
(which is the normal procedure for making a formula prenex) 
while always chosing a variable with minimal $\altfi$. 
Among the prenex normal forms of $\phi$, 
this $\phi'$ is a sensible prenex normal form anyway, 
in that it does not introduce unneccessary alternation. 
Observe that $\gfi=G_{\phi'}$. 
Now by Lemma~\ref{lem:replacements} 
we have $\fotw(\phi)=\fotw(\phi')$ 
while the choice of $\phi'$ 
implies $\tw(\gfi,\altfi)=\tw(G_{\phi'},\alte_{\phi'})$. 

By virtue of Theorem~\ref{theo:fotw-ew} it suffices 
to prove $\ew(\gfi,\ad)\leq\ew(\gfi,\altfi)$. 
We show that for all formulae $\phi$ in prenex normal form, all $k$, 
and all elimination orderings of $(\gfi,\altfi)$ of width $k$, 
there is an elimination ordering of $(\gfi,\ad)$ of width $k$. 
Let $\phi=Q_{x_1}x_1\ldots Q_{x_n}x_n\theta$ 
where $\theta$ is quantifier free. 

For any ordering $y_1,\ldots,y_n$ of $\{x_1,\ldots,x_n\}$, 
a \emph{$\phi$-fault} is a pair $1\leq i<i'\leq n$ 
such that $\ad(y_i)>\ad(y_{i'})$. 
Let us assume that $\phi$ and $y_1,\ldots,y_n$ 
form a counterexample with a minimal number of $\phi$-faults, 
that is $y_1,\ldots,y_n$ is an elimination ordering of $(\gfi,\altfi)$, 
$(\gfi,\ad)$ has no elimination ordering of equal width, 
and the number of $\phi$-faults is minimal 
for all choices of $\phi$ and $y_1,\ldots,y_n$. 
If this number of $\phi$-faults is $0$, 
then $y_1,\ldots,y_n$ also is an elimination ordering of $(\gfi,\ad)$, 
contradicting the counterexample property. 
Hence there is some $\phi$-fault. 

Let $k$ be the width of $y_1,\ldots,y_n$ with respect to $\gfi$. 
Let $\psi:=Q_{y_1}y_1\ldots Q_{y_n}y_n\theta$. 
As $y_1,\ldots,y_n$ is obtained from $x_1,\ldots,x_n$ 
only by rearranging variables within quantifier blocks, 
we can obtain $\psi$ from $\phi$ by a sequence of replacements 
as in Part~2 of Lemma~\ref{lem:replacements}. 
Hence that Lemma implies $\gfi=G_{\psi}$ and $\ad=\dd_{\psi}$. 
Also, $\altfi=\alte_{\psi}$,
so in particular $y_1,\ldots,y_n$ 
is an elimination ordering of $(\psi,\alte_{\psi})$ of width $k$. 
By Definition of $\psi$, we have $y_i\leq_{\psi}y_{i'}$ if and only if $i\leq i'$. 
Furthermore, $\phi$-faults and $\psi$-faults of $y_1,\ldots,y_n$ coincide. 

As there is a $\psi$-fault, 
there also is one which concerns two subsequent variables, 
that is for some $i$ we have $\dd_{\psi}(y_i)>\dd_{\psi}(y_{i+1})$. 
Let $z_1,\ldots,z_n$ be the sequence 
$y_1,\ldots,y_{i-1},$ $y_{i+1},$ $y_i,$ $y_{i+2},\ldots,y_n$, 
i.e.~$y_i$ and $y_{i+1}$ change places. 
Further, let $\chi:=Q_{z_1}z_1\ldots Q_{z_n}z_n\theta$. 
Obviously, $z_1,\ldots,z_n$ 
is an elimination ordering of $(G_{\chi},\alte_{\chi})$. 
We claim that $y_i\not\preceq_{\psi}y_{i+1}$ 
and that $\{y_i,y_{i+1}\}$ is no edge of $G_{\psi,i+1}$, 
where this graph is as in the definition of elimination width 
with respect to the sequence $y_1,\ldots,y_n$. 
These claims are proved later, let us first show how to make use of them. 
The fact $G_{\chi}=G_{\psi}$ and the non-edge between $y_i$ and $y_{i+1}$ 
in $G_{\psi,i+1}$ imply that the width of $z_1,\ldots,z_n$ 
with respect to $\chi$ and $\psi$ is also $k$. 
{}From $y_i\not\preceq_{\psi}y_{i+1}$ it is easy to see 
(and somewhat implicit in the proof of Lemma~\ref{lem:replacements}) 
that $\dd_{\chi}=\dd_{\psi}$. 
In particular, $\psi$-faults and $\chi$-faults coincide. 
By construction, $z_1,\ldots,z_n$ 
has one such fault less than $y_1,\ldots,y_n$, 
so by minimality of the counterexample 
there is some elimination ordering of $(G_{\chi},\dd_{\chi})$ of width $k$. 
As $G_{\chi}=G_{\psi}=G_{\phi}$ and $\dd_{\chi}=\dd_{\psi}=\dd_{\phi}$, 
it is also one of $\phi$ of equal width, 
in contradiction to $\phi,k$ being a counterexample. 

Now for the claims. 
First, assume for contradiction that $y_i\preceq_{\psi}y_{i+1}$. 
As $y_i\not=y_{i+1}$, this is not due to \refl. 
As $y_{i+1}$ is the $\leq_{\psi}$-successor of $y_i$ 
and $(\preceq_{\psi})\subseteq(\leq_{\psi})$, 
there can be no intermediate variable, 
hence $y_i\preceq_{\psi}y_{i+1}$ is due to \alt. 
But then $Q_{y_i}\not=Q_{y_{i+1}}$, 
so $\dd_{\psi}(y_{i+1})\geq\dd_{\psi}(y)+1$ 
in contradiction to $i,i+1$ being a $\psi$-fault. 

For the second claim assume, again for contradiction, 
that $y_i$ forms an edge with $y_{i+1}$ in $G_{\psi,i+1}$, 
i.e.~in $G_{\psi}$ there is a path from $y_i$ to $y_{i+1}$ 
such that all internal vertices of that path 
are of the form $y_j$ with $j>i+1$. 
We view the path as an entanglement chain $y_i=v_0,\ldots,v_n=y_{i+1}$. 
As the elimination ordering is $\leq_{\psi}$, 
we have $y_i,y_{i+1}\leq_{\psi}v_j$ for all $0<j<n$, 
that is the chain is hanging. 
If $Q_{y_i}\not=Q_{y_{i+1}}$, 
then the chain is nice by Lemma~\ref{lem:bloedes zeug 2}. 
In particular, $y_i\preceq_{\psi}y_{i+1}$, contradicting the previous claim. 
Now for the case $Q_{y_i}=Q_{y_{i+1}}$. 
Then $\dd_{\psi}(y_{i+1})<\dd_{\psi}(y_i)$ implies that there is some $w\in\var(\psi)$ 
such that $\dd_{\psi}(w)=\dd_{\psi}(y_i)-1$ and $w\preceq_{\psi}y_i$. 
It follows that $Q_w\not=Q_{y_i}$. 
The fact $w\preceq_{\psi}y_i$ is not due to \refl{} because $Q_w\not=Q_{y_i}$. 
It is also not due to \trans: 
otherwise let $u$ be the intermediate variable. 
Then $Q_u=Q_{y_i}$ or $Q_u=Q_w$, without loss of generality the latter. 
Then $w\preceq_{\psi}u$ is due to \trans{} 
and further unfolding \trans{} until \alt{} is applicable 
eventually yields some $u'$ 
such that $w\preceq_{\psi}u'\preceq_{\psi}u\preceq_{\psi}y_i$ 
and $Q_w\not=Q_{u'}\not=Q_u\not=Q_{y_i}$. 
Then $\dd_{\psi}(w)+1=\dd_{\psi}(y_i)\geq\dd_{\psi}(w)+3$, a contradiction. 
Hence $w\preceq_{\psi}y_i$ is due to \alt, 
so there is some corresponding entanglement chain. 
Prepending it to the one we have 
yields a chain $w=v_{-m},\ldots,v_{-1},v_0=y_i,v_1,\ldots,v_n=y_{i+1}$. 
Furthermore this chain satisfies the first three conditions 
of Lemma~\ref{lem:bloedes zeug 1} (with respect to $\psi$). 
By virtue of Lemma~\ref{lem:bloedes zeug 2}, 
the other two conditions are implied. 
Hence Lemma~\ref{lem:bloedes zeug 1} yields $w\preceq_{\psi}y_{i+1}$, 
so $\dd_{\psi}(y_{i+1})\geq\dd_{\psi}(w)+1=\dd_{\psi}(y_i)$, 
contradicting $\dd_{\psi}(y_{i+1})<\dd_{\psi}(y_i)$. 
\qed

The following example shows that the difference in the opposite
direction can be unbounded.

\begin{exa}\label{ex:fotw-ad}
For $n>0$, 
recall the formula $\phi_n$ from Example~\ref{exa:ent.: dep. on reordering}. 
We have seen that $\dd_{\phi_n}(x_i)=\alte_{\phi_n}(x_i)=1=\dd_{\phi_n}(z)$ 
for $i\leq n$, 
and $\dd_{\phi_n}(y)=\alte_{\phi_n}(y)=2$, 
whereas $\alte_{\phi_n}(z)=3$. 
Figure~\ref{fig:fotw-ad} shows the formula graph of $\phi_n$ 
together with an $\dd_{\phi_n}$-stratified tree decomposition of width $1$ of $\phi_n$.
We actually have $\fotw(\phi_n)= 1$. 
On the other hand, it is easy to see 
that every $\alte_{\phi_n}$-stratified tree decomposition of $\phi_n$ 
has a piece $\var(\phi_n)$ or $\var(\phi_n)\setminus\{y\}$, 
and hence $\tw(G_{\phi_n},\alte_{\phi_n})=n$. 
\end{exa}

\def\vertexnodes{\tikzstyle{every node}=[fill=black,circle,scale=0.5]}
\def\linecloud#1{
\draw[color=black,cap=round,line width=3.1mm] #1;
\draw[color=white,cap=round,line width=2.8mm] #1;
\draw[color=white,style=nearly transparent,cap=round,line width=2.8mm] #1;
}
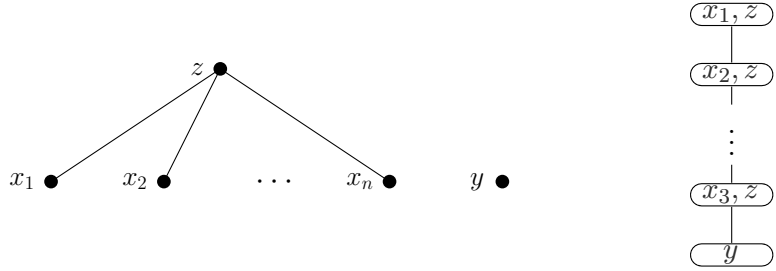
\begin{figure}
\begin{tikzpicture}[yshift=5,baseline,scale=1.5]
\draw (2,0) node[draw=none]{\large{$\cdots$}};
\vertexnodes
\draw (0,0) node[label=left:\huge{$x_1$}]{} -- 
(1.5,1) node[label=left:\huge{$z$}]{} -- 
(1,0) node[label=left:\huge{$x_2$}]{}; 
\draw (1.5,1) -- (3,0) node[label=left:\huge{$x_n$}]{};
\draw (4,0) node[label=left:\huge{$y$}]{};

\end{tikzpicture}$\quad$$\quad$$\quad$$\quad$
\begin{tikzpicture}[baseline,scale=0.8]
\linecloud{(0,3) node{\color{black}$\quad\quad x_1,z$} -- (1,3) node{}};
\linecloud{(0,2) node{\color{black}$\quad\quad x_2,z$} -- (1,2) node{}};
\draw (0.5,1) node[draw=none]{\large{$\vdots$}};
\linecloud{(0,0) node{\color{black}$\quad\quad x_3,z$} -- (1,0) node{}};
\linecloud{(0,-1) node{\color{black}$\quad\quad y$} -- (1,-1) node{}};
\draw(0.5,2.8) -- ++(0cm,-0.6cm);
\draw(0.5,1.8) -- ++(0cm,-0.3cm);
\draw(0.5,0.5) -- ++(0cm,-0.3cm);
\draw(0.5,-0.2) -- ++(0cm,-0.6cm);
\end{tikzpicture}\caption{Example~\ref{ex:fotw-ad}: the formula graph of $\phi_n$ and
an $\dd_{\phi_n}$-stratified tree decomposition of width $1$ of $\phi_n$.}
\label{fig:fotw-ad}
\end{figure}

\section{Computing stratified tree decompositions}\label{sec:computing}

In this section, 
we show that computing stratified tree decompositions of optimal width 
is fixed-parameter tractable, 
where the parameter is the stratified tree-width. 
In fact, the running time is essentially linear 
for bounded stratified tree-width. 
In the next section, we will use the algorithm developed here 
as a first step for formula evaluation. 

\begin{defi}
Let $G=(V,E)$ be a graph and $d:V\to\mathbb N$. 
We say that $(G,d)$ is \emph{normalized}, 
if for $n:=\left|V\right|$ 
we have $V=\{1,\ldots,n\}$ and $d(v)\leq n$ for all $v\in V$. 
\end{defi}

We obtain a linear time algorithm only for normalized inputs. 
In the following, let $(G,d)$ with $G=(V,E)$ be normalized. 
We set $d_{\max}:=\max_{v\in V}d(v)$ 
and for any integer $i$ we let $X_i:=\{v\in V\mid d(v)=i\}$. 
Let $N_G(C)$ denote the set of neighbours of $C$ in $V\setminus C$. 

\begin{defi}
For $0\leq i\leq d_{\max}$, 
let $G^{(i)}$ be the graph with vertex set $V$, 
such that two vertices $x,y$ form an edge in $G^{(i)}$, 
if in $G$ there is a path from $x$ to $y$ 
with all internal vertices in $\bigcup_{j>\max(i,d(x),d(y))}X_j$. 
\end{defi}

In particular, every edge of $G$ is present in all $G^{(i)}$ 
by virtue of a path without any internal vertices. 
For $G^{(d_{\max})}$, we can use only such paths, so $G^{(d_{\max})}=G$. 
Observe the similarity of the graphs $G^{(i)}$ 
to the graphs $G_j$ in the definition of elimination orderings: 
in both cases the connectivity through some vertices 
is redirected to hold immediately between those vertices thus connected. 
The main differences are granularity 
(the number of graphs is $d_{\max}+1$, respectively $|V|+1$) 
and the fact that in the definition of $G^{(i)}$, no vertices are deleted. 
Indeed, we have 
$G^{(i)}[\bigcup\limits_{j\leq i}X_j] = G_{|\bigcup\limits_{j\leq i}X_j|}$. 
This also explains our interest in the $G^{(i)}$. 
We are (implicitly) looking 
for elimination orderings for $G$ which respect $d$. 
Such elimination orderings keep the $X_i$ intact, 
so the $G^{(i)}$ will (up to vertex deletion) 
occur as the $G_j$ at the boundaries between the different $X_i$. 

\begin{defi}
The \emph{component tree} of $(G,d)$ 
is a rooted tree with nodes $t$ 
labelled by two subsets of $V$, denoted by $C_t$ and $D_t$. 
For the root $r$, we let $C_r := V$. 
For a node $t$ at level $i$ (where the root has level $0$), 
we let $D_t := C_t \cap \bigcup_{j\leq i}X_j$. 
For each (nonempty) connected component $C$ of $G[C_t\setminus D_t]$, 
node $t$ has a child $t_C$ and we let $C_{t_C} := C\cup N_G(C)$. 
Further, we let $D^1_t:=D_t\cap X_i$ and $D^2_t:=D_t\setminus D^1_t$. 
\end{defi}

\begin{lem}\label{lem:klebe-cliquen}
Let $G$ be a graph and $d\colon V(G)\to\mathbb N$. 
Let $t,u$ be nodes of the component tree of $(G,d)$, 
where $u$ is a child of $t$. Then
\begin{enumerate}[\em(1)]
	\item $D_t\cap D_u=D_t\cap C_u$.
	\item Let $(T,B)$ be a $d$-stratified tree decomposition of $G$.
		Then some piece of $(T,B)$ covers $D_t\cap D_u$.
\end{enumerate} 
\end{lem}

\proof 
Let $i$ be the depth of $t$ in the component tree and let $C$ be the
connected component of $G[C_t\setminus D_t]$ 
such that $C_u=C\cup N_{G}(C)$.

1) It suffices to prove $D_t\cap D_u\supseteq D_t\cap C_u$.  If $x\in D_t\cap
C_u$, then $x\in D_t\subseteq \bigcup_{0\leq j\leq i}X_j$ and hence $x\in
C_u\cap \bigcup_{0\leq j\leq i+1}X_j=D_u$.

2) By Fact~\ref{fact:cliques} 
  it suffices to show that any pair of distinct vertices $x,y\in D_t\cap D_u$
occurs together in some piece of $(T,B)$.  As $x,y\in D_t$, we have $x,y\not\in
C$, so $x,y\in N_{G}(C)$. 
Since $C$ is a connected component, there is a path $P$
from $x$ to $y$ with all internal vertices in $C$.  Hence $d(z)>d(x)$ and
$d(z)>d(y)$ for all internal vertices $z$ of $P$. 

Let $x=x_0,x_1,\ldots,x_n=y$ be the path $P$.  If $n=1$, then $G$
already contains the edge $\{x,y\}$, which hence is covered by $(T,B)$.
Otherwise, the set of tree nodes covering vertices from $P\setminus\{x,y\}$
induces a nonempty connected subtree $T'$ in $T$.  Let $s$ be the
$\leq_T$-minimal node of $T'$ and let $1<i<n$ be such that $s=t_{x_i}$. 
As $\{x,x_1\}$ is an edge of $G$, 
the vertex $x$ is covered by some node $t'$ of $T'$. 
By definition of $s$ and $t_x$, we have $s\leq_Tt'$ and $t_x\leq_Tt'$. 
Hence, both $t_x$ and $s$ lie on the unique path from the root of $T$
to $t'$ and thus they are comparable by $\leq_T$. 
As $t_{x_i}=s<_Tt_x$ would contradict $d$-stratifiedness, 
$t_x\leq_Ts\leq_Tt'$ follows. 
Using (TD3), we conclude that $s$ covers $x$. 
Analogously, $s$ covers $y$, so $x$ and $y$ occur together in $B_s$. 
\qed

While the component tree is useful for mentally addressing 
the task of computing a stratified tree decomposition, 
we cannot afford to actually compute it: 
its size is superlinear and we want to achieve a linear running time. 
The algorithm instead works with a modified variant. 
First, we can do without the $C_t$s, 
so the algorithm only computes the tree itself and the $D_t$s. 
This projection is necessary to obtain a linear size, 
as is the following second modification: 
in case for some node $t$ at level $i>0$ we have $D_t\cap X_i=\emptyset$, 
then we omit the node $t$, 
making its only child instead a child of the parent of $t$. 
We term $t$ a \emph{dropped node}. 
Whenever we talk about the level of a node in the modified component tree, 
we always refer to the level the node had prior to dropping any nodes. 
As a third modification, for each node $t$, say at level $i$, 
we store $G^{(i-1)}[D_t]$ alongside $D_t$. 
For technical reasons 
which will become clear in the proof of Lemma~\ref{lem:comp-tree}, 
we allow the encodings of these $G^{(i-1)}[D_t]$ to have multiedges. 
To keep the notation simple, 
let us fix the convention that $|E|$ for some edge set $E$ with multiedges 
denotes the sum of multiplicities of edges from $E$. 

\begin{lem}\label{lem:comp-tree}
We can compute a modified component tree of a normalized $(G,d)$ 
in time $\OO{|V|\cdot\tw(G,d)^2}$. 
\end{lem}

\proof
Let $k:=\tw(G,d)$. 
The modified component tree is computed in a bottom-up fashion. 
For this observe, that the various $C$ 
used in the definition of the component tree at level $i$ 
are the connected components of $G[\bigcup\limits_{j\geq i}X_j]$. 
Hence each $C_t$ for $t$ at level $i$ is $C\cup N_G(C)$ for such a $C$. 
In order to determine the corresponding $D_t$, we do not need the full graph: 
it is irrelevant which edges  
between vertices of $\bigcup\limits_{j>i}X_j$ are present; 
it suffices to know 
what connectivity these induce on $\bigcup\limits_{j\leq i}X_j$. 
This information is present in $G^{(i)}$. 
More precisely, we can transform the defining equality 
$D_t=(C\cup N_G(C))\cap\bigcup\limits_{j\leq i}X_j$, 
where $C$ is some connected component of $G[\bigcup\limits_{j\geq i}X_j]$, 
into the equality $D_t=D\cup(N_{G^{(i)}}(D) \cap \bigcup\limits_{j<i}X_j)$, 
where $D$ is the corresponding connected component of $G^{(i)}[X_i]$. 

Recall that, as $(G,d)$ is normalized, 
the vertices of $G$ are $1,\ldots,n$, and $d_{\max}\leq n$, where $n=|V|$. 
More precisely, we assume that $G$ is given as an array of adjacency lists 
and that $d$ is given as an array of $d$'s values, 
where both arrays are indexed with vertices. 
Using bucket sort, we compute the sets $X_0,\ldots,X_{d_{\max}}$ 
in time $\OO{n+d_{\max}}$, which is $\OO n$ thanks to $d_{\max}\leq n$. 

For the bottom-up run, the algorithm uses a loop $i=d_{\max},\ldots,1$. 
As an invariant, at the beginning of run $i$ it has the following data: 
\begin{iteMize}{$\bullet$}
\item The graph $G^{(i)}$ (possibly with multiedges). 
\item An array of lists of subtrees of the modified component tree. 
  Overall, the lists contain all subtrees rooted at level $i+1$. 
  For all $0\leq j\leq d_{\max}$, the list at entry $j$ contains the subtrees which, 
  after dropping nodes, end up at level $j$. 

  For $j>0$, 
  we also store an element from $D_t\cap X_{j-1}$ alongside each subtree, 
  where $t$ is the root of the subtree.\smallskip 
\end{iteMize}

\noindent The graph is initialized to $G^{(d_{\max})}=G$. 
This can be done in zero time, because $G$ is not needed any more. 
The array is initialized with empty lists, 
using time $\OO{d_{\max}}$ and thus $\OO n$. 
For a single run of the loop, 
we first do a depth-first search through $G^{(i)}[X_i]$. 
More precisely, the search is in $G^{(i)}$, 
it uses all elements from $X_i$ as entry points, 
and it terminates recursion in elements from other $X_j$. 
In this way, the connected components $D$ of $G^{(i)}[X_i]$ are found, 
and we also directly obtain 
the corresponding $D\cup (N_{G^{(i)}}(D) \cap \bigcup\limits_{j<i}X_j)$, 
that is some new $D_t$. 
During the search, we do some more things, 
which only increase the running time by a constant factor: 
for each such $D_t$, we generate the tree node $t$ 
and we label each vertex of $D_t$ with $t$. 
The label persists only until the next $i$. 
Also, for each such $t$, while constructing $D_t$, 
we also compute the sets $D^1_t$ and $D^2_t$, 
and we start constructing $G^{(i-1)}[D_t]$ 
by building a graph with vertex set $D_t$ 
and all edges from $G^{(i)}[D_t]$ incident to (at least one vertex of) $D^1_t$. 
Furthermore, 
we determine the maximal $j$, such that $D_t^2\cap X_j\not=\emptyset$ 
and we remember some $x\in D_t^2\cap X_j$. 
The running time of the search 
is $\OO{|X_i|+|\{e\in E(G^{(i)}) | e\cap X_i\not=\emptyset\}|}$. 
The second part, the number (respecting multiplicities) 
of edges of $G^{(i)}$ incident with $X_i$, 
is bounded by the number of edges of $G^{(0)}$ incident with $X_i$. 
Summing over all $i$, the total running time of all searches 
is then $\OO{n+|E(G^{(0)})|}$. 

Next, we update $G^{(i)}$ to $G^{(i-1)}$ 
by introducing, for each new $t$, 
edges between each two vertices from $D^2_t$. 
We also do this in the graph stored at node $t$, 
which thus becomes $G^{(i-1)}[D_t]$ as needed. 
We do not check whether the edges were already present 
(because we do not have enough time to do so), 
hence multiedges may be introduced. 
The running time is $\OO{|D_t|+\ell}$, 
where $\ell$ is the number of edges introduced. 
Summing over all iterations, 
the first part $|D_t|$ is bounded in the same way 
as the running time of the depth-first searches. 
The second part $\ell$ is bounded by $\binom k2$, where $k:=\tw(G,d)+1$: 
as $D_t\cap\bigcup\limits_{j<i}X_j=D_t\cap D_u$ for the parent $u$ of $t$, 
we can conclude $|D^2_t|=|D_t\cap\bigcup\limits_{j<i}X_j|\leq k$ 
from Lemma~\ref{lem:klebe-cliquen},~2. 
Hence in total, the second part induces a running time of $\OO{m\cdot k^2}$, 
where $m$ is the size of the tree. 
As this is the only place, where we add edges to $G^{(i)}$, 
we can also bound $|E(G^{(0)})|$ by $|E(G)|+m\cdot\binom k2$. 

We already have all the nodes for level $i$. 
We only need to connect them to the nodes of level $i+1$ 
(in case $i<d_{\max}$). 
For this, for each tree in the list at the array entry $i+1$, 
say with root $u$, 
we look at its element from $D_u\cap X_i$, say $y$, 
then look at the label of $y$ generated above 
(as $y\in X_i$, we did consider $y$ in the depth-first search), 
say $t$, and make $u$ a child of $t$. 
As this is done at most once for each tree node, 
the total running time of doing this throughout the loop 
is linear in the size of the tree. 
Last, for all new $t$, we recall $j$ and $x$ determined above, 
and we add the tree rooted at $t$ together with $x$ 
to the list at array entry $j+1$. 

This concludes the description of the loop. 
After it has finished, we generate the root node $r$, 
set $D_r:=X_0$ and make all trees in the list at array entry $0$ 
children of $r$. 
The total running time of computing the modified component tree 
is thus $\OO{n+|E(G)|+m\cdot k^2}$, 
where, again, $m$ is the number of nodes of the modified component tree. 
For the second term we have $|E(G)|\leq n\cdot\tw(G) \leq n\cdot k$
by Fact~\ref{fact:number-of-edges}. 
Now, let us estimate $m$. 
Each non-root node $t$ of the modified component tree 
is a non-root node of the original component tree, say at level $i>0$, 
satisfying $D_t\cap X_i\not=\emptyset$. 
Hence, there is an element $x_t\in D_t\cap X_i$. 
Let $u$ be the parent of $t$ in the original component tree. 
Then $x_t$ belongs to the connected component $C$ of $C_u\setminus D_u$, 
such that $C_t=C\cup N_G(C)$. 
For different $t$ at level $i$, 
the $x_t$ come from different connected components, 
so there can be at most $|X_i|$ such $t$. 
In total, $m\leq n$, so the running time is $\OO{n\cdot k^2}$. 
\qed

\begin{rem}\label{rem:klebecliquen}
As a corollary to the previous proof, 
let us observe that $G^{(0)}$ is obtained from $G$ 
by turning all $D_t^2$ into cliques. 
\end{rem}

\begin{thm}\label{theo:computing-decompositions}
There is an algorithm that, given a normalized $(G,d)$, 
computes a $d$-stratified tree decomposition of $G$ of minimum width 
in time $\OO{|V(G)|\cdot 2^{\poly(\tw(G,d))}}$. 
\end{thm}

\proof[Proof] 
Observe that, if $u$ is a child of $t$ in the component tree, 
then $D_t\cap D_u=D_u^2$. 
Then Lemma~\ref{lem:klebe-cliquen},~2 and Remark~\ref{rem:klebecliquen} 
imply that $d$-stratified tree decompositions of $G$ 
coincide with those of $G^{(0)}$. 
Hence it suffices to compute a $d$-stratified
tree decomposition for $G^{(0)}$ of minimum width. 
This is achieved in four phases. 
First, a modified component tree is computed as in Lemma~\ref{lem:comp-tree}. 

In the second phase, for each node $t$ of the component tree, an optimal tree
decomposition of $G^{(0)}[D_t]$ is computed. 
Observe, that, if $i$ is the level of $t$, 
then $G^{(0)}[D_t]$ coincides with $G^{(i-1)}[D_t]$, 
which we have stored at node $t$. 
Hence, for a single $t$, using Bodlaender's algorithm \cite{Bodlaender96} 
(which works well even in the presence of multiedges), 
the tree decomposition can be obtained 
in time $\|G^{(i-1)}[D_t]\|\cdot 2^{\poly(k)}$. 
Observe, that the sum of all $\|G^{(i-1)}[D_t]\|$ 
is bounded by the encoding size of the modified component tree, 
which is bounded by the time it took to compute it. 
Hence, overall this phase takes time $n\cdot 2^{\poly(k)}$. 

In the third phase, for each node $t$ of the component tree 
and each parent or child $u$ of $t$, 
we determine a node $x_{t,u}$ of the tree decomposition of $G^{(0)}[D_t]$, 
whose piece fully contains $D_t\cap D_u$. 
Observe, that $D_t\cap D_u$ coincides with $D^2_t$ or with $D^2_u$, 
depending on whether $u$ is a parent or a child of $t$. 
Hence, $D_t\cap D_u$ is a clique in $G^{(0)}$, 
and then also in $G^{(0)}[D_t]$, so there must be some $x_{t,u}$ as above. 
We find $x_{t,u}$ by first reading $D^2_t$ respectively $D^2_u$ 
from the modified component tree, 
and then checking each element from each piece of the tree decomposition 
against this set $D_t\cap D_u$, 
until some adequate $x_{t,u}$ is found. 
Even without sophisticated search structures, 
$\OO{\ell\cdot |D_t\cap D_u| \leq \ell\cdot k}$ time suffices, 
where $\ell$ is the size of the tree decomposition. 
Overall, this phase takes time $\OO{n\cdot k^3}$. 

In the fourth phase, we use a bottom-up recursion 
to construct, for all nodes $t$, 
an optimal tree decompositions of $G^{(0)}[C_t]$. 
It will happen to extend the above tree decomposition of $G^{(0)}[D_t]$. 
The recursion works as follows: 
we cycle through the children of $t$. 
By virtue of bottom-up-ness, for each such child $u$, 
we already know an optimal tree decomposition of $G^{(0)}[C_u]$. 
As it contains the known decomposition of $G^{(0)}[D_u]$, 
it also contains the node $x_{u,t}$. 
Hence we connect it to the known decomposition of $G^{(0)}[D_t]$ 
by adding an edge between $x_{u,t}$ and $x_{t,u}$. 
The result still is a tree decomposition,
because the intersection $D_t\cap D_u$ separates $D_t$ from $C_u$
(Lemma~\ref{lem:klebe-cliquen},~1). 
It is an optimal tree decomposition, 
because its width is the maximum 
of the widths of the participating tree decompositions,
which were optimal for their respective subgraphs. 
In the end, the recursion yields an optimal tree 
decomposition $(T,B)$ of $G^{(0)}[C_r]=G^{(0)}$. 
It takes time $\OO m$. 

Last, pick some node $s$ from the tree decomposition of $G^{(0)}[D_r]$. 
As $B_s \subseteq D_r \subseteq X_0$, 
choosing $s$ as the root of $(T,B)$ makes it $d$-stratified. 
This concludes the proof of Theorem~\ref{theo:computing-decompositions}.
\qed

As each $(G,d)$ can be normalized in time $\OO{n\log n}$, 
a similar statement holds for arbitrary $(G,d)$, 
albeit not with linear running time. 
Theorem~\ref{theo:computing-decompositions} implies 
that we can efficiently decide whether a formula
$\phi$ satisfies $\fotw(\phi)\leq k$: 

\begin{cor}\label{cor:computing-decompositions}
There is an algorithm that, given a formula $\phi\in\ml$, 
computes a tree decomposition of $\phi$ of minimum width 
in time $\poly(\|\phi\|)+\left|\var(\phi)\right|\cdot 2^{\poly(\fotw(\phi))}$. 
\end{cor}

\proof
Given $\phi$, it is first turned into a straight formula. 
Then, $G_{\phi}$ and $\ad$ are computed and normalized in polynomial time. 
The last step is a call 
to the algorithm from Theorem~\ref{theo:computing-decompositions}. 
\qed

\section{Query evaluation on bounded first order tree-width}\label{section:mc}

This section contains the second main result: 
evaluating formulae of bounded tree-width 
is fixed parameter tractable with parameter the length of the formula.
Moreover, we show that evaluating quantified constraint formulae of bounded
first order tree-width can be done in polynomial time. 
This is stronger than 
Chen and Dalmau's result~\cite{chedal05} for quantified constraint formulae of bounded 
elimination-width: as we will see in Section~\ref{section:related-notions}, 
bounded elimination-width (i.e.\ bounded \emph{tree-width} 
in~\cite{chedal05}) implies bounded first order tree-width, but there
are classes of quantified constraint formulae with unbounded elimination-width, that
have bounded first order tree-width.

\begin{defi}
Let $k\geq 0$ be an integer. 
The fragment $\fok$ of $\ml$ consists 
of those (not necessarily straight) formulae $\phi$ 
such that $\left|\var(\phi)\right|\leq k$. 
In contrast, let $\fokm k$ be the fragment of formulae $\phi$ of $\ml$ 
such that all subformulae of $\phi$ have at most $k$ free variables. 
\end{defi}

\begin{rem}\label{rem:foks und fokm}
Obviously, $\fok\subseteq\fokm k$. 
On the other hand, formulae from $\fokm k$ can be turned 
into equivalent formulae from $\fok$ by renaming bound variables. 
The algorithm runs in polynomial time, and in linear time for fixed $k$. 
\qed\end{rem}

By Remark~\ref{rem:foks und fokm}, 
we can use $\fok$ and $\fokm k$ interchangeably. 
It will be more convenient to work with the latter. 

\begin{rem}
For all $k$, 
the question whether a given formula is equivalent to an $\fok$ formula 
is undecidable. 
\end{rem}

The statement is folklore, but we provide a proof for completeness' sake. 

\proof
We reduce from satisfiability of $\ml$. 
By introducing a new unary predicate $U$ 
and relativizing all quantifiers to $U$, 
satisfiability reduces to the question, 
whether a given formula from $\ml$ is satisfiable by an infinite structure. 
By the theorem of L\"owenheim and Skolem, this in turn is equivalent 
to satisfiability by a countable infinite structure. 
Now let $\phi$ be some fixed formula 
such that $\phi$ does not have finite models, 
and such that $\phi$ is not equivalent to any $\fok$-formula. 
Then a given formula $\psi$ is unsatisfiable by infinite structures, 
if and only if $\phi\wedge\psi'$ is equivalent to an $\fok$ formula, 
where $\psi'$ is obtained from $\psi$ 
by renaming all symbols 
(constant symbols, relation symbols, and free variables) 
to be disjoint from all symbols of $\phi$. 
For the \lq{}only if\rq{} part, 
unsatisfiability by infinite structures of $\psi$ 
implies the same for $\psi'$. 
Then, $\phi\wedge\psi'$ is unsatisfiable and hence equivalent 
to $Pc\wedge\neg Pc$ from $\foks0$. 
For the \lq{}if\rq{} part, 
assume that $\psi$ is satisfiable 
and that $\phi\wedge\psi'$ is equivalent to $\chi\in\fok$. 
We obtain $\chi'$ from $\chi$ 
by renaming all symbols which do not occur in $\phi$ 
to be disjoint from all symbols of $\psi'$. 
By choice of $\chi$ we have $\chi\models\phi$ and then $\chi'\models\phi$. 
The converse would contradict 
that $\phi$ is not equivalent to any $\fok$-formula, 
so there is some interpretation $\mathcal I_1$ 
such that $\mathcal I_1\models\phi\wedge\neg\chi'$. 
As $\phi$ does not have any finite models, 
the domain of $\mathcal I_1$ is infinite, and by L\"owenheim-Skolem 
we can assume without loss of generality that it is countable. 
On the other hand, satisfiability of $\psi$ implies that of $\psi'$, 
say by an interpretation $\mathcal I_2$ with countable infinite domain. 
By virtue of all the renaming done above, 
$\mathcal I_1$ and $\mathcal I_2$ do not have any symbols in common. 
Furthermore, their domains are of the same cardinality. 
Hence there is an interpretation $\mathcal I$ which extends both 
(up to isomorphisms), 
that is we have $\mathcal I\models\phi\wedge\psi'$ 
but $\mathcal I\not\models\chi$, 
a contradiction to the assumed equivalence. 
\qed

As a consequence, there is no computable width parameter 
such that width-$k$ captures all formulae 
logically equivalent to a formula of $\fok$. 
For any width parameter, only formulae 
which are \lq{}syntactically close\rq{} to a formula of $\fok$ are captured, 
for varying values of \lq{}syntactically close\rq{}. 

The following example shows that first order tree-width indeed does not
capture equivalence to $\foks2$.

\begin{exa}
	Let $n>2$. $\phi_n=\psi_n\vee\chi$, where 
	$\chi\in \foks2$ and
	$\fotw(\psi_n)=n$, but $\psi_n$ is unsatisfiable.  Then
	$\phi_n\equiv\chi\in$ $\foks2$, but $\fotw(\phi_n)=n$.
\end{exa}

First order sentences of tree-width at most $k-1$ have the same expressive
power as $\fokm k$ (and by Remark~\ref{rem:foks und fokm} hence as $\fok$). 
More generally we have the following.

\begin{thm}\label{theo:fotw-fok}
Let $k\ge 0$. 
\begin{enumerate}[\em(1)]
\item For any formula $\phi$ with $\fotw(\phi)\le k-1$ there is a formula
  $\psi\in\fokm k$ with $\phi\equiv\psi$ which is computable from $\phi$. 
\item Any formula $\psi\in\fokm k$ satisfies $\fotw(\psi)\le k-1$.
\end{enumerate}
\end{thm}

\proof
\setcounter{clam}0
2. By Theorem~\ref{thm:fotw und ad} it suffices to show 
  that $\tw(G_{\psi},\alte_{\psi})\leq k-1$. 
  Without loss of generality, $\psi$ is already straight. 
  Let $T$ be the syntax tree of $\psi$. 
  For a node $t$, say $t$ corresponds to the subformula $\chi$ of $\psi$, 
  let $B_t$ consist of the free variables of $\chi$. 
  As $\psi\in\fokm k$, the width of $(T,B)$ is at most $k-1$. 
(TD2) holds for $(T,B)$, because edges arise from atoms and atoms are
among the subformulae $\chi$ considered in the definition of $(T,B)$. 
Then (TD1) follows, because $\psi$ is straight. 
(TD3) holds, because for all variables $x$, 
the set $\{t\in T\mid x\in B_t\}$ union of all paths 
from (nodes corresponding to) atoms in which $x$ occurs 
to (the node corresponding to) the scope of $x$. 
  It is easy to see that $(T,B)$ is $\alte_{\psi}$-stratified. 

1. Given $\phi$, 
we use the algorithm of Corollary~\ref{cor:computing-decompositions} 
to compute an $\ad$-stratified tree decomposition $(T,B)$ 
of width $\fotw(\phi)\le k-1$ for $\gfi$. 

Intuitively, we will iteratively replace subformulae of $\phi$ 
by equivalent $\fokm k$-formulae, 
until we obtain an $\fokm k$ formula $\psi$ equivalent to $\phi$.
The replacement is done along a tree decomposition $(T,B)$ of $\phi$ of width
at most $k-1$. For our purposes
it is more convenient only to work in leaves of the decomposition, so in every
iteration we restrict the decomposition to the part of the formula that still
has to be transformed into an $\fokm k$ formula. 
In doing so, 
we treat the subformulae $\phi'$ of $\phi$ already in $\fokm k$ as new atoms, 
defined on the variables $\free(\phi')$. 
For this we make sure that these new atoms 
are covered in some piece of the remaining part of the tree decomposition. 

More precisely, we describe an iterative algorithm. 
In every iteration, we are given a formula $\phi'$ in xenerp normal form, 
a tree decomposition $(T',B')$ of $\phi'$ of width at most $k-1$, 
and a second order substitution $S$, 
substituting relation symbols of $\phi'$ 
by $\fokm k$ formulae with the appropriate number of free variables, 
such that we have $\phi'S\equiv\phi$.
We will need to extend the syntax of formulae: 
we allow any monotone Boolean function as a single Boolean connective. 
Of course, this extended syntax still allows for negation normal form, 
and we still assume that all formulae are in this normal form. 
This syntax extension carries over 
to the definition of $\phi_{[X]}$ and hence of $\ad$.

We start by letting $\phi'$ be some xenerp normal form of $\phi$, 
$(T',B')=(T,B)$, and $S=\emptyset$. 
Now in every step we do the following. 

1. If $\phi'$ is quantifier free, then $\var(\phi')=\free(\phi)$. 
As $|\free(\phi)|\leq\fotw(\phi)+1\leq k$ 
(recall, that $\free(\phi)$ forms a clique in $\gfi$), 
we have $\phi'\in\fokm k$, so the algorithm stops with output $\phi'S$. 
More precisely, $\phi'S$ may still use our additional Boolean connectives, 
but it is trivial to eliminate these without leaving $\fokm k$ 
(but in general making the formula non-straight). 

2. Otherwise, if $(T',B')$ has a leaf $\ell$ with parent $t$ satisfying
$B_{\ell}\subseteq B_t$, then we remove $\ell$ and $B_{\ell}$ from $(T',B')$,
keeping $\phi'$ and $S$.

3. If neither 1 nor 2 apply, 
then we choose a variable $x\in \var(\phi')$ 
as in the following claim.

\begin{clam}
There exists a bound variable $x\in \var(\phi')$
with $\dd_{\phi'}(x)$ maximum, 
such that $x$ appears in exactly one piece $B'_{\ell}$ of $(T',B')$, 
and $\ell$ is a leaf. 
\end{clam}

\proof
This claim is a variant of Claim~\ref{claim:ex x d-max} 
in the proof of Theorem~\ref{theo:fotw-ew}, 
and in can be shown in the same way. 
Instead of smallness of the decomposition, 
we use the fact that Case~2 does not apply. 
Thus we obtain a variable $x$ with $\dd_{\phi'}(x)$ maximum 
and a leaf $\ell$ such that $x$ appears only in $B_{\ell}$. 
As Case~1 does not apply, there are bound variables. 
As $\dd_{\phi'}(x)$ is maximum, $x$ is a bound variable. 
\cqed

Without loss of generality, suppose that $Q_x=\exists$. 
Let $\psi$ be the scope of $x$ in $\phi'$ 
and let $V$ be the set of variables quantified in $\psi$. 
We partition $V$ into the subsets $V_1$ and $V_2$, 
where $V_1$ contains the variables with which $x$ is entangled in $\phi'$. 
As $\phi'$ is xenerp, $\psi$ is 
$\phi'_x$ preceded by some quantifications of variables from $V_1$. 
By maximality of $\dd_{\phi'}(x)$, we have $Q_y=\exists$ for all $y\in V_1$. 

\begin{clam}\label{claim:V2 </= V1}
No variable $y\in V_1$ is quantified in the scope of some variable $z\in V_2$. 
\end{clam}

\proof
Otherwise, $x$ occurs in $\phi'_y$ (because $x$ and $y$ are entangled), 
$\phi'_y$ is a subformula of $\phi'_z$ 
(because $\phi'$ is xenerp and $z\leq_{\phi'}y$), 
and $z$ occurs in $\phi'_z$ which is a subformula of $\phi'_x$. 
Hence $x$ and $z$ are entangled, contradicting $z\in V_2$.
Thus the claim holds.
\cqed

\begin{clam}\label{claim:x preceq-max}
$x$ is $\preceq_{\phi'}$-maximal. 
\end{clam}

\proof
Otherwise there is some $y\not=x$ such that $x\preceq_{\phi'}y$. 
In the inductive definition of $\preceq_{\phi'}$, 
the pair $(x,y)$ is not introduced by \refl. 
If it is introduced by \trans, say with intermediate variable $z$, 
we can replace $y$ by $z$, 
so we can eventually assume that $x\preceq_{\phi'}y$ is due to \alt. 
In particular $Q_x\not=Q_y$. 
Thus $\dd_{\phi'}(y)>\dd_{\phi'}(x)$, 
contradicting $\dd_{\phi'}$-maximality of $x$. 
\cqed

In the same way we can show
that there are no $\preceq_{\phi'}$-relationships among $V_1$. 

Let $\psi'$ be obtained from $\psi$ 
by removing all quantifications for variables from $V_1$. 
Letting $\exists V_1$ denote the sequence of all these quantifications 
in arbitrary order, 
$\exists V_1\psi'$ is obtained from $\psi$ 
by a sequence of replacements 
as in Lemma~\ref{lem:replacements} parts 1~and~2. 
In particular, $\exists x\psi$ is equivalent to $\exists x\exists V_1\psi'$. 

As all variables quantified in $\psi'$ are from $V_2$ 
and their scopes in $\psi'$ are the same as in $\psi$, 
we have that $x$ does not occur in any quantified subformula of $\psi'$. 
Hence $\psi'$ is a positive Boolean combination 
of atoms, of negated atoms, and of subformulae in which $x$ does not occur. 
By choosing the subformulae in which $x$ does not occur maximal, 
$\psi'$ is a positive Boolean combination 
of atoms using $x$, of negated atoms using $x$, 
and of maximal subformulae of $\psi'$ in which $x$ does not occur. 
We transform this Boolean combination 
into disjunctive normal form $\bigvee_{i\in I} \bigwedge_{j\in J_i}L_j$ 
with literals $L_j$. 
Here each $L_j$ is an atom using $x$, a negated atom using $x$, 
or a maximal subformula of $\psi'$ in which $x$ does not occur. 
Then 
\[
  \exists x\psi
  \equiv \exists x\exists V_1\psi'
  \equiv \exists x\exists V_1\bigvee_{i\in I} \bigwedge_{j\in J_i}L_j
  \equiv \exists V_1\bigvee_{i\in I}\exists x\bigwedge_{j\in J_i}L_j.
\]
For $i\in I$ let $J_i^-\subseteq J_i$ be 
the subset of indices $j$ such that $x$ does not occur in $L_j$ 
and let $J_i^+ =J_i\setminus J_i^-$. 
For convenience we let $J^+:=\bigcup\limits_{i\in I}J^+_i$ 
and $J^-:=\bigcup\limits_{i\in I}J^-_i$. 
Now 
\[
  \exists x \psi
  \equiv \exists V_1\bigvee_{i\in I}\exists x\bigwedge_{j\in J_i}L_j
  \equiv \exists V_1\bigvee_{i\in I}\big(\bigwedge_{j\in J_i^-}L_j
    \wedge\exists x\bigwedge_{j\in J_i^+}L_j\big).
\]
Let $j\in J^+$. 
As $(T',B')$ is a tree decomposition of $\phi'$, 
(the atom underlying) $L_j$ is covered by some piece of $(T',B')$. 
Since $x$ is a variable of $L_j$ and $x$ occurs in $B'_{\ell}$ only, 
$L_j$ is covered in $B'_{\ell}$. 
Hence for $i\in I$ and $\psi_i^x:=\exists x\bigwedge_{j\in J_i^+}L_j$ 
we have $\var(\psi_i^x)\subseteq B'_{\ell}$. 
Let $x\bar y$ be an enumeration of $B'_{\ell}$, 
and for all $i\in I$ 
let $A_i^x$ be a new relation symbol of arity $|B'_{\ell}|-1\leq k-1$. 
Replace $\exists x\psi$ in $\phi'$ by $\exists V_1\psi''$, where 
\[
  \psi'':=
    \bigvee_{i\in I}\big(\bigwedge_{j\in J_i^-}L_j\wedge A_i^x\bar y \big)
\]
and let $\phi''$ be the formula thus obtained from $\phi'$. 
Let $S_x$ be the substitution which replaces every $A_i^x\bar y$ 
by $\psi_i^x$, respectively. 
Clearly, $\phi' \equiv \phi''S_x$, so by setting $S':=S_x S$ 
we have $\phi \equiv \phi'S \equiv \phi''S'$. 
More precisely, in the definition of $\psi''$, 
we use a single positive Boolean connective 
(as allowed by the above syntax extension) 
for the entire disjunctive normal form. 
This Boolean connective has only one input for each $L_j$ used. 
In particular, we retain a single quantifier 
per variable quantified in $\psi$, that is we retain straightness. 

We obtain a tree decomposition $(T'',B'')$ for $\phi''$ by letting
$T'':=T'$ and removing $x$ from $B'_{\ell}$. 
This tree decomposition still covers all 
(edges of $G_{\phi''}$ created by) atoms of $\phi''$ 
which also occur in $\phi'$, because these do not use $x$. 
The new atoms $A_i^x\bar y$ are covered by $B''_{\ell}=B'_{\ell}\setminus\{x\}$. 
This shows (TD2).
(TD1) and (TD3) are inherited from $(T',B')$. 

In the rest of the proof we show that $(T'',B'')$ 
is $\dd_{\phi''}$-stratified. 

\begin{clam}\label{claim:entanglement phi' vs phi''}
Let $y$ be a bound variable and $Z$ a set of bound variables 
such that $x\not=y$ and $x\not\in Z$. 
Then $y$ occurs in $\phi''_{[Z]}$ 
if and only if $y$ occurs in $\phi'_{[Z]}$ 
or $\phi'_{[Z]}$ is a subformula of $\phi'_x$ 
but not of any $L_j$ with $j\in J^-$ 
and $y$ occurs in $\phi'_x$. 
\end{clam}

\proof
Observe that $\phi'_{[Z]}$ is generated 
already by at most two occurrences of variables from $Z$. 
If one of these occurences is outside of $\psi$, 
then $\phi'_{[Z]}$ is disjoint from $\psi$ and $\phi'_x$, 
or $\psi$ and $\phi'_x$ are proper subformulae of $\phi'_{[Z]}$. 
In either case we have to show 
that the same variables $y\not=x$ occur in $\phi'_{[Z]}$ as in $\phi''_{[Z]}$. 
This follows, because in the first case $\phi'_{[Z]}=\phi''_{[Z]}$ 
whereas in the second case $\phi''_{[Z]}$ is obtained from $\phi'_{[Z]}$ 
by replacing the subformula $\exists x\psi$ with $ \exists V_1\psi''$. 
It remains to consider the case 
where all (generating) occurrences of variables from $Z$ in $\phi'$ 
are in $\psi$. 
Let $j_1,j_2$ be such 
that the generating occurrences are in $L_{j_1}$ and $L_{j_2}$. 
If $j_1=j_2\in J^-$, 
then $\phi'_{[Z]}$ is a subformula of $L_{j_1}$. 
Furthermore $\phi''_{[Z]}=\phi'_{[Z]}$, 
so the same variables $y$ occur in these formulae. 
If any generating variable is from $B'_{\ell}$, 
then it occurs in all atoms $A_i^x\bar y$, 
so $\phi''_{[Y]}=\psi''$. 
In the last case, $j_1,j_2\in J^-$ and $j_1\not=j_2$. 
Again, $\phi''_{[Y]}=\psi''$. 
In both cases there is no single $j\in J^-$ 
such that $\phi'_{[Z]}$ is a subformula of $L_j$, 
so we need to show that the same variables $y$ 
occur in $\phi''_{[Z]}$ as in $\phi'_x$ 
(including the special case of occurrence 
in the subformula  $\phi'_{[Z]}$ of $\phi'_x$). 
This follows because $\phi''_{[Z]}=\psi''$ 
and $\psi''$ uses the same variables as $\phi'_x$. 
This shows the claim. 
\cqed

For arbitrary variables $y,z$ different from $x$, 
the two following claims show 
that $y\preceq_{\phi'}z$ if and only if $y\preceq_{\phi''}z$. 

\begin{clam}
Let $y,z$ be variables different from $x$. 
Then $y\preceq_{\phi'}z$ implies $y\preceq_{\phi''}z$. 
\end{clam}

\proof
First, consider the following modification of $\preceq_{\phi'}$. 
Relax, in the definition of entanglement 
with respect to $\trianglelefteq$ and $\phi'$, 
the conditions of the form \lq$v$ occurs in $\phi'_{[w\trianglelefteq]}$\rq{} 
by \lq$v$ occurs in $\phi'_{[w\trianglelefteq]}$ 
or in $\phi''_{[(w\trianglelefteq)\setminus\{x\}]}$\rq, 
with the convention that no variable occurs in $\phi''_{[\emptyset]}$. 
Let $\preceq$ denote the modified relation. 
It is clear that $y\preceq_{\phi'}z$ implies $y\preceq z$. 
By induction on the definition of $\preceq$ 
it is easy to see that $x$ also is $\preceq$-maximal. 

In order to transform derivations for $\preceq$ 
into derivations for $\preceq_{\phi''}$, 
we start by normalizing them. 
First, we may assume that no entanglement chain 
used in any application of \alt{} repeats elements. 
Also, we may assume that \refl{} and \trans{} 
are always applied as early as possible. 
In particular, at each application of \alt, 
the current approximation for $\preceq$ is reflexive and transitive. 
The third normalization concerns applications of \alt, 
say with chain $v_0,\ldots,v_n$ 
and with $\trianglelefteq$ as the current approximation of $\preceq$. 
Such an application is normalized, if for entanglement, 
instead of the sets $v_i\trianglelefteq$, 
already the sets $(v_i\trianglelefteq)\setminus\{x\}$ suffice 
(unless $v_i=x$, in which case $(v_i\trianglelefteq)=\{x\}$ 
due to maximality of $x$). 
Now we claim that this normalization is always possible. 
We prove the claim by induction on derivations for $\preceq$ 
which are already in the first two normal forms. 
Only the \alt{} step is nontrivial, 
so consider an application of \alt{} with chain $v_0,\ldots,v_n$ 
and approximation $\trianglelefteq$ for $\preceq$. 
By the inductive hypothesis, 
all pairs from $\trianglelefteq$ can be derived in a normalized way. 
If our application of \alt{} is not already normalized, 
then there is some $0\leq i\leq n$ such that $x\not=v_i\trianglelefteq x$. 
After unfolding \trans, 
we obtain some $w$ such that $v_i\trianglelefteq w$ 
and such that $w\trianglelefteq x$ holds due to \alt{} (possibly $w=v_i$). 
Let $(\trianglelefteq')\subseteq(\trianglelefteq)$ 
be the approximation of $\preceq$ pertaining to this application of \alt. 
Then there is another chain of entanglements 
ending with $u,x$, such that $w\trianglelefteq'u$ or $x\trianglelefteq'u$. 
Due to the normalizations from the inductive hypothesis, 
$x$ already occurs in $\phi'_{[(u\trianglelefteq')\setminus\{x\}]}$ 
or in $\phi''_{[(u\trianglelefteq')\setminus\{x\}]}$. 
As $x$ does not occur in $\phi''$ at all, the former must be the case. 
As $x\trianglelefteq'u$ would contradict $\preceq$-maximality of $x$, 
we have $w\trianglelefteq'u$. 
Transitivity gives $v_i\trianglelefteq u$. 
Now consider one of the up to two neighbours of $v_i$ 
in the entanglement chain, 
without loss of generality we pick $v_{i+1}$. 
We distinguish the cases $v_{i+1}=x$ and $v_{i+1}\not=x$. 
In the first case, 
we already know 
that $x$ occurs in $\phi'_{[(u\trianglelefteq')\setminus\{x\}]}$. 
{}From $v_i\trianglelefteq u$ and transitivity we conclude 
$(u\trianglelefteq')\setminus\{x\}\subseteq(v_i\trianglelefteq)\setminus\{x\}$, 
so $x$ occurs in $\phi'_{[(v_i\trianglelefteq)\setminus\{x\}]}$ as needed. 
Now for the case that $v_{i+1}\not=x$. 
As $v_{i+1}$ occurs in $\phi'_{[v_i\trianglelefteq]}$ 
or in $\phi''_{[(v_i\trianglelefteq)\setminus\{x\}]}$, 
it is easy to see that $v_{i+1}$ 
also occurs in $\phi''_{[(v_i\trianglelefteq)\setminus\{x\}\cup\{u\}]}$. 
But as $v_i\trianglelefteq u$, 
this is $\phi''_{[(v_i\trianglelefteq)\setminus\{x\}]}$. 
In a last normalization step, we also eliminate $x$ 
from occurrences in entanglement chains in derivations. 
Hence assume that $v,x,w$ is part of such a chain, 
again with $\trianglelefteq$ as approximation of $\preceq$. 
As $x$ is $\preceq$-maximal, we have $(x\trianglelefteq)=\{x\}$, 
hence $v$ occurs in $\phi'_{[\{x\}]}$ and then also in $\psi''$. 
As $x$ occurs in $\phi'_{[(w\trianglelefteq)\setminus\{x\}]}$ 
(recall the previous normalization), 
we conclude that $\psi''$ is a subformula 
of $\phi''_{[(w\trianglelefteq)\setminus\{x\}]}$. 
Thus $v$ occurs in $\phi''_{[(w\trianglelefteq)\setminus\{x\}]}$ 
and similarly $w$ occurs in $\phi''_{[(v\trianglelefteq)\setminus\{x\}]}$, 
so $x$ can be omitted from the chain. 
After this normalization, we can assume 
that $x$ occurs in applications of \alt{} only as an endpoint of the chain. 

Now assume some counterexample $y,z$ to the claim. Then $y\preceq z$. 
Let $y,z$ be derivation-minimal 
with respect to normalized derivations for $\preceq$. 
Then $y\not=z$ because $\preceq_{\phi''}$ is reflexive. 
Next assume that $y\preceq z$ is due to transitivity, 
say with intermediate variable $v$. 
If $v=x$ then $x\preceq z\not=x$ contradicting $\preceq$-maximality of $x$. 
Thus $v\not=x$, 
so $y\preceq_{\phi''}v\preceq_{\phi''}z$ by minimality of the counterexample. 
$y\preceq_{\phi''}z$ follows. 
Finally for \alt, 
let us assume some chain $y=v_0,\ldots,v_n=z$ 
of entanglements in the sense of $\preceq$, 
such that $y\preceq v_i$ or $z\preceq v_i$ for all $0\leq i<n$. 
As always, denote the approximation of $\preceq$ by $\trianglelefteq$. 
By normalization of derivations, no $v_i$ equals $x$ 
and $x$ is not needed for entanglement. 
{}From derivation-minimality we conclude 
$y\preceq_{\phi''}v_i$ or $z\preceq_{\phi''}v_i$, 
and that $(v_i\trianglelefteq)\setminus\{x\}\subseteq(v_i\preceq_{\phi''})$ 
for all $0\leq i\leq n$. 
By Claim~\ref{claim:entanglement phi' vs phi''}, 
the occurrence of $v_{i+1}$ in $\phi'_{[(v_i\trianglelefteq)\setminus\{x\}]}$ 
or $\phi''_{[(v_i\trianglelefteq)\setminus\{x\}]}$ 
implies occurrence of $v_{i+1}$ in $\phi''_{[v_i\preceq_{\phi''}]}$. 
Similarly, $v_i$ occurs in $\phi''_{[v_{i+1}\preceq_{\phi''}]}$. 
Thus, the entanglement chain is also such a one in $\phi''$. 
Also $y\leq_{\phi'}z$ implies $y\leq_{\phi''}z$ 
so we conclude $y\preceq_{\phi''}z$. 
\cqed

Now for the converse. 

\begin{clam}
Let $y,z$ be variables different from $x$. 
Then $y\preceq_{\phi''}z$ implies $y\preceq_{\phi'}z$. 
\end{clam}

\proof
We work by induction on $\preceq_{\phi''}$. 
The cases \refl{} and \trans{} are immediate. 
For \alt{} let $\trianglelefteq$ be some approximation of $\preceq_{\phi''}$, 
assume $y\leq_{\phi''}z$ and $Q_y\not=Q_z$ 
and let $y=w_0,\ldots,w_n=z$ be given 
such that for all $0\leq i<n$ we have 
$y\preceq_{\phi''}w_i$ or $z\preceq_{\phi''}w_i$, 
occurrence of $w_i$ in $\phi''_{[w_{i+1}\trianglelefteq]}$, 
and occurrence of $w_{i+1}$ in $\phi''_{[w_i\trianglelefteq]}$. 
The inductive hypotheses are 
$y\preceq_{\phi'}w_i$ or $z\preceq_{\phi'}w_i$, 
and $(\trianglelefteq)\subseteq(\preceq_{\phi'})$. 
As $\leq_{\phi'}$ and $\leq_{\phi''}$ differ only among $V_1$ 
and we have $Q_y\not=Q_z$, we conclude $y\leq_{\phi'}z$. 
If furthermore all occurrences also hold in $\phi'$, then we are done. 
So assume that there is some $i$ 
such that $w_i$ is not entangled with $w_{i+1}$ in $\phi'$, 
that is $w_i$ does not occur in $\phi'_{w_{i+1}}$ 
or $w_{i+1}$ does not occur in $\phi'_{w_i}$. 
Without loss of generality let us assume the former. 
Claim~\ref{claim:entanglement phi' vs phi''} then implies 
that $w_i$ occurs in $\phi'_x$ and that $\phi'_{w_{i+1}}$ 
is a subformula of $\phi'_x$ but not of any $L_j$ with $j\in J^-$. 
By maximality of these $L_j$ among the subformulae not containing $x$, 
this implies that $x$ occurs in $\phi'_{w_{i+1}}$. 
Of course $w_{i+1}$ occurs in $\phi'_{w_{i+1}}$ and thus in $\phi'_x$. 
Hence $x$ and $w_{i+1}$ are entangled in $\phi'$. 
Next let us show that also $w_i$ and $x$ are entangled. 
If $w_{i+1}$ does not occur in $\phi'_{w_i}$, then we can proceed as above, 
so we may assume that $w_{i+1}$ does occur in $\phi'_{w_i}$. 
Then $w_i$ and $w_{i+1}$ are comparable by $\leq_{\phi'}$. 
If $w_{i+1}\leq_{\phi'}w_i$, then xenerp normal form of $\phi'$ 
implies that $w_i$ occurs in $\phi'_{w_{i+1}}$ 
in contradiction to our assumption. 
Hence $w_i\leq_{\phi'}w_{i+1}$ 
and by xenerp we have that $\phi'_{w_{i+1}}$ is a subformula of $\phi'_{w_i}$. 
In particular $x$ occurs in $\phi'_{w_i}$ and we already know the converse. 
In this way, whenever a link of the $\phi''$-chain does not hold in $\phi'$, 
we can insert $x$ to obtain a longer valid chain. 
So far, validity only means that each link is an entanglement in $\phi'$. 

It remains to show that $y\preceq_{\phi'}x$ or $z\preceq_{\phi'}x$, 
if $x$ needed to be inserted. 
Let $v$ and $v'$ be the neighbours of (one occurence of) $x$ in the chain. 
If $v$ and $v'$ both are in $\{y,z\}$, 
then, because $x$ is only inserted between non-entangled variables, 
one of them is $y$ and the other one is $z$. 
As $Q_y\not=Q_z$, we have $Q_v\not=Q_x$ or $Q_{v'}\not=Q_x$, 
without loss of generality the former. 
As $x$ and $v$ are entangled in $\phi'$ by the above, 
they are comparable by $\leq_{\phi'}$. 
Thus by \alt{} we either have $x\preceq_{\phi'}v$, 
contradicting $\preceq_{\phi'}$-maximality of $x$, 
or $v\preceq_{\phi'}x$, in which case we are done because $v\in\{y,z\}$. 
Now for the case that one of $v$ and $v'$ is not from $\{y,z\}$, 
without loss of generality $y\not=v\not=z$. 
Still, $v$ is entangled with $x$. 
Furthermore, $v$ is an original element of the chain, 
so $y\preceq_{\phi'}v$ or $z\preceq_{\phi'}v$. 
Without loss of generality let us assume the latter 
and recall that $z\not=v$. 
After unfolding, in the derivation of $z\preceq_{\phi'}v$, 
some applications of \trans, 
we obtain some $u$ such that $z\preceq_{\phi'}u\preceq_{\phi'}v$ 
and $u\preceq_{\phi'}v$ is due to \alt. 
Accordingly, let $u=t_0,\ldots,t_k=v$ be a witnessing entanglement chain. 
If $v\leq_{\phi'}x$, then we can extend this chain by $x$ 
to obtain (together with the fact $u\preceq_{\phi'}v$) 
a witnessing chain for $u\preceq_{\phi'}x$ 
which implies $z\preceq_{\phi'}x$ using \trans. 
It remains to consider the case $v\not\leq_{\phi'}x$. 
But as $v,x$ are entangled, they are comparable by $\leq_{\phi'}$, 
so $x\leq_{\phi'}v$ and thus $v\in V_1$ 
implying $Q_v=Q_x$ and then $Q_u\not=Q_x$. 
As $u\preceq_{\phi'}v$ we have $u\leq_{\phi'}v$, 
so $x$ and $u$ are comparable by $\leq_{\phi'}$. 
$x\leq_{\phi'}u$ would imply $u\in V_2$ (since $Q_u\not=Q_x$), 
contradicting Claim~\ref{claim:V2 </= V1}. 
Hence $u\leq_{\phi'}x$ so the chain $t_0,\ldots,t_k$ 
can be extended by $x$ to show $u\preceq_{\phi'}x$ 
and then $z\preceq_{\phi'}x$. 
\cqed

Hence $\preceq_{\phi'}$ and $\preceq_{\phi''}$ coincide outside of $x$. 
As $x$ is $\preceq_{\phi'}$-maximal, 
this implies that $\dd_{\phi'}$ and $\dd_{\phi''}$ coincide outside of $x$. 
Hence $\dd_{\phi'}$-stratification of $(T',B')$ 
implies $\dd_{\phi''}$-stratification of $(T'',B'')$. 
Finally, we turn $\phi''$ into xenerp normal form. 

It is easy to see that the algorithm terminates: 
the second case decreases the size of the tree decomposition 
while it leaves the set of variables intact. 
The third case leaves the tree intact and eliminates one variable. 
As both the tree and the set of variables are finite, 
eventually the first case must trigger and the algorithm stops. 
This concludes the proof of Theorem~\ref{theo:fotw-fok}
\qed

It is well-known that first order query evaluation for $\fok$ 
can be done in time $n^{k+\OO1}$, see~\cite{vardi95}. 

\begin{cor}\label{cor:fo-eval}
	Evaluating queries of bounded first order tree-width is fixed
	parameter tractable with parameter the length of the formula. 
	
	More precisely, given a finite structure $\mathcal A$ 
and a formula $\phi\in\ml$, 
there is an algorithm that computes $\phi(\mathcal A)$ in time 
$\|\mathcal A\|^{\fotw(\phi)+\OO1}f(\left|\varphi\right|)$
for some computable function $f$.
\qed\end{cor}

Here the function $f$ is basically the running time we need for translating 
the formula $\phi$ with $\fotw(\phi)\leq k-1$
into an $\fok$ formula. It is $q$-times exponential, 
where $q$ is the alternation depth of $\phi$. 
The exponentiations arise from converting some subformulae 
into disjunctive or conjunctive normal form. 
In special cases where this step is not needed, the running time is much lower. 

\begin{cor}\label{cor:mc in ptime}
Evaluating formulae without disjunctions and of bounded first order tree-width 
can be done in polynomial time. 
\end{cor}

\proof
Let $\phi$ be a formula without disjunctions such that $\fotw(\phi)<k$. 
Let us recall the algorithm 
underlying the proof of Theorem~\ref{theo:fotw-fok},~1. 
The argument for termination (at the very end of the proof) 
actually shows that the main loop is iterated only a quadratic number of times. 
Case~1 of the main loop is only executed once, 
and it is polynomial in the data then present. 
Case~2 basically requires a search through the current tree, 
which is a subtree of the original one. 
It remains to show that Case~3 runs in polynomial time 
and to bound the way it increases the size of the data. 
In fact, all individual steps of Case~3 
run in polynomial time even for general $\phi$, 
with the exception of turning $\psi'$ into disjunctive normal form. 
But as our $\phi$ does not contain any disjunctions, 
the same holds for $\phi'$ and $\psi'$, 
so $\psi'$ already is in disjunctive normal form and there is nothing to do. 
Formally, the index set $I$ only contains one element 
and (up to reordering) we have $\psi'=\psi_-\wedge\psi_+$, 
where $\psi_-:=\bigwedge\limits_{j\in J^-}L_j$ 
and $\psi_+:=\bigwedge\limits_{j\in J^+}L_j$. 

Let us now bound the data increase. 
The data consist of the tree decomposition (which only becomes smaller), 
the formula $\phi'$, and the substitution $S$. 
It is easier to consider, instead of $\phi'$ and $S$, the formula $\phi'S$. 
The size difference between the two 
stems only from a collection of pairs $(x,i)$, 
where $x$ is a variable and $i$ is an index from the respective $I$. 
As each such $I$ is a singleton, there are at most $|\var(\phi)|$ such pairs. 
With respect to $\phi'S$, all that happens in Case~3 
is shifting quantifiers 
and then replacing $\exists x\exists V_1(\psi_-\wedge\psi_+)$ 
by $\exists V_1(\psi_-\wedge\exists x\psi_+)$. 
The size does not change. 

Hence the algorithm from Theorem~\ref{theo:fotw-fok},~1 
runs in polynomial time. 
So does turning the formula from $\fokm k$ into one from $\fok$. 
We conclude with the time $n^{k+\OO1}$ for evaluating the latter. 
\qed

Of course, the same applies to formulae without conjunctions. 
One example of formulae without disjunctions 
are quantified constraint formulae, 
which we define and discuss in detail in Section~\ref{section:related-notions}. 

The fixed parameter tractability from Corollary~\ref{cor:fo-eval} 
does not remain when we make the first order tree-width part of the parameter. 
This is even true for model checking instead of evaluation. 
Instead, the problem becomes $\makemathname{AW}[*]$-hard: 
$\fotw$ is bounded by the length of the formula, 
and the model checking problem for first order logic, 
parameterized by the formula length, 
is $\makemathname{AW}[*]$-complete~\cite{DFT1996}.

\section{Relation to similar notions}\label{section:related-notions}

In this section we show that $\fotw$ and tree-width coincide on conjunctive
queries, while $\fotw$ is more powerful than both elimination-width of 
quantified
constraint formulae and strict tree-width of non-recursive stratified
datalog programs. Finally, we extend the cops and robber game characterizing
tree-width to stratified tree-width and we prove that requiring monotonicity
does not limit the cops.

A \emph{(Boolean) conjunctive query} is a sentence 
$
	\phi=\exists x_1\ldots\exists x_n\psi,
$
where $\psi$ is a conjunction of relational atoms such that
$\var(\psi)=\{x_1,\ldots, x_n\}$. The \emph{tree-width} of a
conjunctive query $\phi$, $\tw(\phi)$, is defined as the tree-width of $\gfi$ 
(see \cite{KolaitisV00}). 
Any conjunctive query $\phi$ satisfies $\ad=1$. Hence the notion of $\fotw$
generalises the notion of tree width of conjunctive queries. 

\begin{rem}\label{rem:CQs}
Any conjunctive query $\phi$ satisfies $\fotw(\phi)=\tw(\phi)$.\qed
\end{rem}


\subsection{Quantified constraint formulae}

A \emph{quantified constraint formula} \cite{chedal05} is a sentence
\[
\phi=Q_1 x_1Q_2 x_2\ldots Q_nx_n\psi,
\]
where $Q_i\in\{\forall,\exists\}$ for $i=1\ldots n$ and 
$\psi$ is a conjunction of relational atoms.
In \cite{chedal05}, Chen and Dalmau introduce the notion of \emph{tree-width}
of a quantified constraint formula and they show that model checking
for quantified constraint formulae of bounded tree-width can be done in polynomial 
time using the $k$-consistency algorithm. Since their notion of tree-width
is defined via an elimination ordering rather than via a decomposition, 
we call it \emph{elimination-width}
instead of \emph{tree-width}. 
We show that $\fotw$ is less than or equal to Chen and Dalmau's elimination-width,
and we give an example of a class of quantified constraint formulae 
having bounded $\fotw$ and unbounded elimination-width.
By Corollary~\ref{cor:mc in ptime}, 
model checking of quantified constraint formulae of bounded $\fotw$ 
can also be done in polynomial time. 

Recall that by Theorem~\ref{theo:fotw-ew}, any graph $G$ and
$d:V(G)\to\mathbb N$ satisfies $\tw(G,d)=\ew(G,d).$
Let $\varphi$ be a quantified constraint formula with formula graph $\gfi$. 
Let $\altfi'$ be the mapping that assigns to a variable $v\in\var(\varphi)$ the
number of quantifier changes occurring before $v$ in the
quantifier prefix of $\varphi$, adding $+1$ 
(note that here we always add $+1$, 
regardless of the first quantifier of $\phi$). 
Then Chen and Dalmau's notion of
elimination-width can be equivalently 
reformulated in our setting as $\ew(\gfi,\altfi')$.  

\begin{lem}\label{lem:fotw <= CDw}
	Let $\varphi$ be a first order  formula. Then 
$
 	\fotw(\varphi) \leq \ew(\gfi,\altfi'). 
$
\end{lem}

\proof 
By Theorem~\ref{theo:fotw-ew}, we may use $\tw$ and $\ew$ interchangeably. 
Then, by Theorem~\ref{thm:fotw und ad}, 
it suffices to show $\tw(\gfi,\altfi)=\tw(\gfi,\altfi')$. 
First, assume that $\phi$ starts with a quantifier. 
If this quantifier is existential, then $\altfi=\altfi'$. 
If it is universal, then $\altfi(x)=\altfi'(x)=0$ for all free variables $x$, 
and $\altfi(x)-1=\altfi'(x)>0$ for all bound variables $x$. 
In both cases, an elimination ordering respects $\altfi$ 
if, and only if, it respects $\altfi'$. 
Hence $\ew(\gfi,\altfi)=\ew(\gfi,\altfi')$ 
and thus $\tw(\gfi,\altfi)=\tw(\gfi,\altfi')$. 

The same holds trivially, when $\phi$ is quantifier free. 

Now for the general case: $\phi$ is a positive boolean combination 
of formulae $\phi_1,\ldots,\phi_n$ 
which are quantifier free or start with quantifiers. 
Let $F$ be the set of free variables of $\phi$. 
We already know 
that $\tw(G_{\phi_i},\alte_{\phi_i})=\tw(G_{\phi_i},\alte'_{\phi_i})$ 
for all $1\leq i\leq n$. 
As the various $\phi_i$ have pairwise distinct bound variables, 
$\gfi-F$ is the disjoint union of the $G_{\phi_i}-F$. 
Then it is easy to see that 
\[ \tw(\gfi,\altfi)
  =\max(|F|-1,\max\limits_{1\leq i\leq n}\tw(G_{\phi_i},\alte_{\phi_i})) \]
and
\[ \tw(\gfi,\altfi')
  =\max(|F|-1,\max\limits_{1\leq i\leq n}\tw(G_{\phi_i},\alte'_{\phi_i})) , \]
from which it follows that $\tw(\gfi,\altfi)=\tw(\gfi,\altfi')$.\qed 

\begin{rem}\label{rem:fotw-CD}
	There exists a class of quantified constraint formulae having first order
	tree-width $1$, where Chen and Dalmau's elimination-width is unbounded: 
	let the class consist of the $\phi_n$ as in Example~\ref{ex:fotw-ad}. Then 
	$\alte'_{\phi_n}=\alte_{\phi_n}$, and hence 
	$\tw(G_{\phi_n},\alte'_{\phi_n})=\ew(G_{\phi_n},\alte'_{\phi_n})=n$, while
	$\fotw(\phi_n)= 1$. 
\qed\end{rem}


\subsection{Non-recursive stratified datalog}
In \cite{flufrigro01}, Flum, Frick and Grohe define strict tree-width of non-recursive
stratified datalog (\nrsd) programs and they show that the evaluation problem
for \nrsd{} programs can be solved in polynomial time on programs of bounded
strict tree-width \cite[Corollary 5.26]{flufrigro01}.  \nrsd{} programs have the same
expressive power as $\ml$ 
and there are simple translations in both directions. 
This allows us to compare their notion
with $\fotw$. We show that if the \nrsd{} program has tree-width at most $k$,
then the corresponding first order formula has $\fotw$ at most $k$, and we
exhibit a class of formulae with bounded $\fotw$, whose corresponding \nrsd{}
programs  have unbounded tree-width.


We assume that the reader is familiar with datalog and we only 
fix our notation, and we refer the reader to~\cite{flufrigro01}
otherwise.
A \emph{datalog rule} $\rho$ \emph{with negation} is an expression 
$
	Qx_1\ldots x_l\leftarrow\bigwedge_{i=1}^n\lambda_i,
$ 
where $Q$ is a relation symbol and $x_1,\ldots ,x_n\in\var(\bigwedge_{i=1}^n\lambda_l)$ 
are pairwise distinct variables, and the $\lambda_i$ are literals
($i=1,\ldots ,n$). $Qx_1\ldots x_l$ is called the \emph{head} of
$\rho$, and $\bigwedge_{i=1}^n\lambda_i$ is called the \emph{body} of
$\rho$. To define the semantics, let $\mathcal A$ be a structure
whose vocabulary contains all the relation symbols occurring in the body
of $\rho$. Let $\bar y$ be a tuple that consists of all variables 
of $\var(\bigwedge_{i=1}^n\lambda_i)\setminus\{x_1,\ldots,x_l\}$,
and let 
$\phi_{\rho}(x_1\ldots x_l)=\exists\bar y\bigwedge_{1\leq i\leq n}\lambda_i$.
We let $\rho(\mathcal A):=\phi_{\rho}(\mathcal A)$.
A non-recursive stratified datalog 
(\nrsd{}) program is a sequence $\Pi=(\Pi^1,\ldots,\Pi^n)$
of non-recursive datalog programs $\Pi^i$ (called the strata of $\Pi$) 
as defined in~\cite{flufrigro01}. 
We denote the intentional vocabulary of $\Pi$ by $\Int(\Pi)$
and the extensional vocabulary of $\Pi$ by $\ext(\Pi)$.

The \emph{strict tree-width} of a datalog rule 
$\rho$ is defined as $\stw(\rho):=\tw(G_{\phi_{\rho}})$, and for 
an \nrsd{} program $\Pi=(\Pi^1,\ldots,\Pi^n)$ the \emph{strict tree-width} of
$\Pi$ is defined as $\stw(\Pi):=\max\{\stw(\rho)\mid\rho\in \bigcup_{i=1}^n\Pi^i\}$.%
\footnote{In \cite{flufrigro01}, the term \emph{strict} tree-width refers to the fact 
that tree decompositions are required
to cover all variables in the head of a datalog rule together in some piece.}
The following is proved in \cite{flufrigro01}, Corollary 5.26 (2).
\begin{thm}[Flum, Frick, Grohe]
	For fixed integer $k>0$, the evaluation problem for \nrsd{} programs
	of strict tree-width at most $k$ can be solved in polynomial time.
\qed\end{thm}

It is well known that a query is definable in first order logic if and only
if it is \nrsd{} definable. Actually, an \nrsd{} program $\Pi_Q$ (i.e.\ an
\nrsd{} program with goal predicate $Q$) defines
an equivalent first order formula $\phi_{\Pi_Q}$ in a natural way, and 
vice versa.
For proving our theorem, we make one direction explicit, 
associating a first order formula to 
an \nrsd{} program as follows. 

Let $\Pi=(\Pi^1,\ldots,\Pi^n)$ be an \nrsd{} program and let $Q\in\Int(\Pi^i)$
for some $i\leq n$. Suppose for all $Q'\in \Int(\Pi)$ occuring in 
$ (\Pi^1,\ldots,\Pi^{i-1})$, the formula $\phi_{\Pi_{Q'}}$ is already defined.
Let
\[\bar \phi_{\Pi_Q}:=
\bigvee_{\rho\in\Pi^i,\; Q\text{ occurs in the head of } \rho}\phi_{\rho},\]
and let
\[\phi_{\Pi_Q}:=
\bigvee_{\rho\in\Pi^i,\; Q\text{ occurs in the head of } \rho}\phi^*_{\rho},\]
where $\phi^*_{\rho}$ is obtained from $\phi_{\rho}$ by recursively replacing 
relation symbols $Q'\in\Int(\Pi)$ occuring in $\phi_{\rho}$ by the
corresponding formula $\bar\phi_{\Pi_{Q'}}$ (i.e.\ $\phi_{\Pi_Q}$ is an 
$\ext(\Pi)$-formula).
Let $\mathcal A$ be an $\ext(\Pi)$-structure. It is easy to see
that we have
$\Pi_Q(\mathcal A)=\phi_{\Pi_Q}(\mathcal A).$

Recall that $\alte'_{\phi}$ 
is the mapping that assigns to a variable $v\in\var(\varphi)$ the
alternation depth of $v$ in $\phi$. 

\begin{thm}\label{app:theo:datalog}
	Any \nrsd{} program $\Pi$ with $Q\in \Int(\Pi)$ satisfies\\ 
	$\fotw(\phi_{\Pi_Q})\leq\tw(G_{\phi_{\Pi}},\alte'_{\phi_{\Pi_Q}})\leq\stw(\Pi)$. 
\end{thm}

\proof 
The first inequality follows from  Theorem~\ref{theo:fotw-ew}
and Lemma~\ref{lem:fotw <= CDw}.

Towards the
second inequality, let
$\Pi=(\Pi^1,\ldots, \Pi^n)$ and let
$k:=\stw(\Pi)=\max\{\stw(\phi_{\rho})\mid \rho\in \bigcup_{i=1}^n\Pi_i\}$.
We prove by induction on the number $n$ of strata of $\Pi$ that all
$Q\in\Int(\Pi)$ satisfy $\tw(G_{\phi_{\Pi_Q}},\alte'_{\phi_{\Pi_Q}})\leq k$.
Let $Q\bar x$ be the head corresponding to $Q$ 
and let $i$ be such that $Q\in\Int(\Pi^i)$. 
Then
$\phi_{\Pi_Q}$ 
has exactly the free variables $\bar x$ and we have 
\[\phi_{\Pi_Q}=
\bigvee_{\rho\in\Pi^i,\; Q\text{ occurs in the head of } \rho}\phi^*_{\rho}.\]


\noindent Suppose all $Q'\in\Int(\Pi)$
occuring in $(\Pi^1,\ldots, \Pi^{i-1})$
satisfy $\tw(\alte'_{\phi_{\Pi_{Q'}}},G_{\phi_{\Pi_{Q'}}})\leq k$.

We may assume that $\bar \phi_{\Pi_Q}$ and $\phi_{\Pi_Q}$
are straight.

First we construct a tree decomposition for 
$\bar \phi_{\Pi_Q}$ as follows.
For every $\rho\in\Pi^i$ with head $Q\bar x$,
we take a tree decomposition 
of width at most $k$ of $G_{\phi_{\rho}}$. Each
of these decompositions has a piece containing 
the variables $\bar x$.
We glue them
together at one new root covering the variables
$\bar x$. Then we orient the edges of the decomposition tree 
away from the root, and we obtain a tree decomposition
$(\bar T,\bar B)$ for $\bar \phi_{\Pi_Q}$ of width at most $k$.
By the inductive 
hypothesis, for every $Q'\in\Int(\Pi)$ such that
$Q'\bar y$ occurs in $\bar \phi_{\Pi_Q}$,
we have an $\alte'_{\phi_{\Pi_{Q'}}}$-stratified tree decomposition $(T^{Q'},B^{Q'})$ of
$\phi_{\Pi_{Q'}}(\bar y)$. By definition, the variables in $\bar y$
are contained in the piece at the root $r$ of $T^{Q'}$. Moreover, $\bar y$
is covered in some piece of $(\bar T,\bar B)$. We choose such a piece $\bar B_t$
and we attach $(T^{Q'},B^{Q'})$
to this piece such that $r$ becomes a new successor of $t$. 
Having done this for all atoms $Q'\bar y$ (with $Q\neq Q'$), 
that occur in $\bar \phi_{\Pi_Q}$, we obtain a tree decomposition $(T,B)$ for 
$G_{\phi_{\Pi_Q}}$ of width at most $k$. We may assume that 
$\phi_{\Pi_Q}$ is in negation normal form. (If not, we transform 
$\phi_{\Pi_Q}$ into negation normal form. Note that this does not change
the formula graph.)
Then, by construction, $(T,B)$ is 
$\alte'_{\phi_{\Pi_Q}}$-stratified.
\qed

The following remark shows that the 
difference can be unbounded in the opposite direction. Moreover, it
shows that the difference between Chen and Dalmau's elimination-width 
and tree-width of \nrsd{} programs can be unbounded.

\begin{rem}\label{rem:fotw-better-datalogtw}
	There is a class $\mathcal C$ of \nrsd{} programs with unbounded strict tree-width,
	such that $\fotw(\phi_{\Pi})=\tw(G_{\phi_{\Pi}},\alte'_{\phi_{\Pi}})=0$ 
for all $\Pi\in\mathcal C$. 
\end{rem}
\proof
For an integer $n>0$
let 
$
\psi_n:=\exists x_1\ldots \exists x_{n-1}\forall x_n
\big(\bigwedge_{i=1}^{n}Px_i\big)$. 
Take $\mathcal C$ to consist of the natural \nrsd{} programs 
which are equivalent to the formulae $\psi_n$, for $n>0$. 
\qed

\subsection{Cops, Robbers and stratified tree-width}
We now introduce the cops and robbers game as defined
in \cite{seytho93}.
Let $G$ be a graph and let $k\geq 0$ be an integer.
The cops and robbers game on $G$
(with game parameter $k$) is played by two players, the
\emph{cop player} and the \emph{robber player}, on the graph $G$. 
The cop player controls $k$ cops and the robber player controls 
the robber.
Both the cops and the robber move on the vertices of $G$.
Some of the cops move to at most $k$ vertices and the robber stands on
a vertex $r$ not occupied by the cops.
In each move, some of the cops fly in helicopters 
to at most $k$ new vertices. 
During the flight, the robber sees which position the 
cops are approaching and before they land she quickly tries to 
escape by running arbitrarily fast along paths of $G$
to a vertex $r'$, 
not being allowed to run through a standing cop.
Hence, if $X\subseteq V(G)$ is the cops' first position,
the robber stands on $r\in V(G)\setminus X$, and
after the flight, the cops occupy the set $Y\subseteq V(G)$,
then the robber can run to any vertex $r'$ within
the connected component of $G\setminus (X\cap Y)$ containing $r$.
The cops win if they 
land a cop via helicopter on the vertex occupied by the robber.
The robber wins if she can always elude capture.
\emph{Winning strategies} are defined in the usual way.
The \emph{cop-width} of $G$, $\cw(G)$, is the minimum number of
cops having a winning strategy on $G$.

A winning strategy for the cops is \emph{monotone}, 
if for all plays played according to the strategy, 
if $X_1,X_2,\ldots$ is the sequence of cop positions, then 
the connected components $R_i$ of $G\setminus X_i$ containing the
robber form a decreasing (with respect to $\subseteq$) 
sequence.
The $R_i$ are called the \emph{robber spaces}. 
The \emph{monotone cop-width} of $G$, $\moncw(G)$, is the minimum number of
cops having a monotone winning strategy on $G$.

\begin{thm}[Seymour, Thomas \cite{seytho93}]\label{theo:seytho}
	Any graph $G$ satisfies $\tw(G)+1=\cw(G)=\moncw(G)$.
\qed\end{thm}

Now let $G$ be a graph and let $d$ be a function $d\colon V(G)\to\mathbb N$.
The \emph{$d$-stratified} cops and robbers game on $G$ 
is played as the cops and robbers game on $G$, 
but in every move the cops have to satisfy the following additional
condition. Intuitively, they can only clear vertices $v$ with $d(v)=i$
after they have cleared all vertices $w$ with $d(w)<i$.
More precisely:
for every move $(X,R)$, where $X\subseteq V(G)$ is the cop position
and $R$ is the robber space, the cops have to make sure that
$\max\{d(x)\mid x\in X\}\leq\min\{d(r)\mid r\in R\}$.
Then $\cw(G,d)$ and $\moncw(G,d)$ are defined
analogously, and for a formula $\phi$ we let
$\cw(\phi):=\cw(\gfi,\ad)$ and $\moncw(\phi):=\moncw(\gfi,\ad)$.

Although proving the following theorem is not very hard, it seems interesting
to know that $\cw(G,d)$ and $\moncw(G,d)$ coincide. In many generalisations
of the cops and robbers game to other settings, the analogous statements  
become false \cite{adl04,Adler07,kreord08}, and it might be helpful to 
explore the borderline.

\begin{thm}\label{theo:games}
	Let $G$ be a graph and $d:V(G)\to\mathbb N$.
	The following statements
	are equivalent:
	\begin{enumerate}[\em(1)]
		\item $\tw(G,d)\leq k-1$, 
		\item $\moncw(G,d)\leq k$, and
		\item $\cw(G,d)\leq k$.
	\end {enumerate}
In particular, any first order formula $\phi$ satisfies
	$\fotw(\phi)+1=\moncw(\phi)=\cw(\phi)$.
\end{thm}

\proof
\setcounter{clam}0
1 $\Rightarrow$ 2: suppose $\tw(G,d)\leq k-1$. Let $(T,B)$ be a 
	$d$-stratified tree decomposition of width at most $k-1$ for $G$.
{}From $(T,B)$, the $k$ cops can read off a monotone winning strategy 
in the usual way, first moving to $B_r$ 
and then following the robber down the tree decomposition into the unique
	direction where the robber space is covered (see e.g.~\cite{Adler08a}).	
Since $(T,B)$ is $d$-stratified, the winning strategy is also $d$-stratified.
	
	2 $\Rightarrow$ 3: any monotone winning strategy is a winning strategy.

	3 $\Rightarrow$ 1: suppose $k$ cops have a winning strategy 
for the $d$-stratified game on $G$.

\begin{clam}
Let $t,u$ be nodes of $(G,d)$'s component tree, where $u$ is a child of $t$.
Let $x,y\in D_t\cap D_u$.
If, while playing against the $k$ cops, 
the robber can move to $x$ and no cop will land on $y$, then
the robber can also move to $y$.
\end{clam}

Let $i$ be the depth of
$t$ in $(G,d)$'s component tree and let $C$ be the connected component of
$G[C_t\setminus D_t]$ with $C_u=C\cup N(C)$. 
As $x,y\in D_t$ we have $x,y\not\in C$, and hence $x,y\in N_{G}(C)$ 
and there is a path from $x$ to $y$ with all internal vertices
		in $C$. Hence $d(z)>i\geq d(x)$ and $d(z)>i\geq d(y)$ for all internal vertices
		of the path. Therefore, as long as the cops play on vertices with $d$ at most $i$, 
		the path is free and the robber can use it. But the cops
      can never move to a vertex with $d>i$ before they have cleared $D_t\cap D_u$ completely,
		so the path from $x$ to $y$ is free whenever the robber can move to $x$. 
\cqed

Recall, from Section~\ref{sec:computing}, 
that the graph $G^{(0)}$ is obtained from $G$ 
by adding all edges between any pair of distinct vertices $x,y\in D_t\cap D_u$ 
for all directed edges $(t,u)$ of the component tree. 
By the claim, any $d$-stratified winning strategy for $k$ cops on $G$ is 
also a $d$-stratified winning strategy on $G^{(0)}$. 
Forgetting $d$, obviously, $k$ cops have a winning strategy on $G^{(0)}$ 
and hence in particular on $G^{(0)}[D_t]$ 
for all nodes $t$ of $(G,d)$'s component tree. 
Thus $\cw(G^{(0)}[D_t])\leq k$ 
and by Theorem \ref{theo:seytho} this implies $\tw(G^{(0)}[D_t])\leq k-1$. 
{}From the tree decompositions of the $G^{(0)}[D_t]$ of width $\leq k$ 
we can now construct tree decompositions for the $G^{(0)}[C_t]$ 
of width $\leq k-1$ in a bottom-up manner 
as in the proof of Theorem \ref{theo:computing-decompositions}. 
For the root $r$ of $(G,d)$'s component tree, 
we have $G^{(0)}=G^{(0)}[C_r]$ and it is easy to see 
that the tree decomposition for $G^{(0)}[C_r]$ of width $\leq k$ 
obtained in this way is $d$-stratified. 
\qed

\section{Conclusion}\label{section:conclusion}

We introduced a notion of tree-width for first order formulae $\phi$,
$\fotw(\phi)$,
generalising tree-width of conjunctive queries and
elimination-width of quantified constraint formulae \cite{chedal05}.
Our notion can also be seen as an adjustment of the notion of tree-width of first order
formulae as defined in \cite{flufrigro01} (which only works for conjunctive
queries with negation).

We proved that computing $\fotw$ 
is fixed-parameter tractable with parameter $\fotw$
(Theorem \ref{theo:computing-decompositions}).
Moreover, we showed that evaluating formulae of $k$-bounded
first order tree-width is fixed-parameter tractable, with
parameter the length of the formula (Theorem \ref{theo:fotw-fok}). 
This is done by first computing a tree decomposition of width at most $k$
for the formula, and then translating
the formula equivalently into a formula of the $k$-variable
fragment $\fok$ of first order logic. It is well-known that
evaluating $\fok$ formulae can be done in polynomial time.
When translating the formula $\phi$ into an equivalent $\fok$
formula, we get a non-elementary explosion in the running time. 

\begin{conj}
When translating a formula $\phi$ satisfying $\fotw(\phi)\leq k$ 
into an equivalent $\fok$
formula, a non-elementary explosion cannot be avoided.
\end{conj}

Moreover, it is still unknown whether the explosion can be avoided 
in parameterized algorithms 
for evaluating queries of bounded first order tree-width. 

We show that first order tree-width can be characterised by other notions
such as elimination-width (Theorem \ref{theo:fotw-ew}), and the minimum number
of cops necessary to catch the robber in the stratified cops and robbers
game, as well as the minimum number of cops necessary in the monotone
version of the game (Theorem \ref{theo:games}). Hence our notion is very natural
and robust. 

Moreover, we showed that $\fotw$ is more powerful than the notion of
elimination-width 
of quantified constraint formulae as defined in \cite{chedal05}:
for quantified constraint formulae, both bounded elimination-width
and bounded $\fotw$ allow for model checking in polynomial time.
We proved that if $\phi$ is a quantified constraint formula, then
$\fotw(\phi)$ is bounded by the elimination-width of $\phi$, and
there are classes of quantified constraint formulae
with bounded $\fotw$ and unbounded elimination-width.

Finally, we showed that $\fotw$ is more powerful than   
tree-width of non-recursive stratified datalog (\nrsd) programs \cite{flufrigro01}.
\nrsd{} programs have the same
expressive power as first order logic, in the sense that \nrsd{} programs correspond
to first order formulae and vice versa.
We showed 
that first-order tree-width of (formula versions of) \nrsd{} programs 
is bounded by the strict tree-width of the programs 
and that there are classes of first order formulae with bounded $\fotw$,
whose corresponding $\nrsd$ programs have unbounded strict tree-width. 

For conjunctive query evaluation, methods more powerful than
bounded tree-width are known. Conjunctive queries of bounded hypertree-width 
\cite{gotleosca02},
bounded fractional hypertree-width \cite{gromar06} 
and bounded (hyper)closure tree-width \cite{Adler08}
yield even larger tractable classes of instances. 
For example, conjunctive queries of bounded hypertree-width
correspond to the $k$-guarded fragment of first order logic
\cite{gotleosca03}, and similar correspondences can be found for the
other invariants. Why not generalise these notions to first order
formulae? By generalising these
notions to first order formulae $\phi$ in the obvious way, a decomposition
of bounded width would not give us an instruction
how to translate $\phi$ into the corresponding guarded
fragment of first order logic (transforming subformulae of $\phi$
into conjunctive normal form as in the proof of
Theorem \ref{theo:fotw-fok}, 1 does not necessarily
yield guarded subformulae).

Nevertheless, generalising these notions to quantified constraint
formulae should indeed yield classes with an efficient query
evaluation, that are strictly larger than classes of quantified constraint
formulae of bounded first order tree-width. It would be interesting to find
the largest fragment of first order formulae 
for which such a generalization is possible. 

{
\bibliographystyle{plain}
\bibliography{fotw}
}

\end{document}